# Direct phase-sensitive identification of a *d*-form factor density wave in underdoped cuprates

K. Fujita[a,b,c,†], M. H. Hamidian[a,b,†], S.D. Edkins[b,d], Chung Koo Kim[a], Y. Kohsaka[e],
M. Azuma[f], M. Takano[g], H. Takagi[c,h,i], H. Eisaki[j], S. Uchida[c], A. Allais[k],
M. J. Lawler[b,l], E.-A. Kim[b], S. Sachdev[k,m] & J. C. Séamus Davis[a,b,d]

- a. CMPMS Department, Brookhaven National Laboratory, Upton, NY 11973, USA.
- b. LASSP, Department of Physics, Cornell University, Ithaca, NY 14853, USA.
- c. Department of Physics, University of Tokyo, Bunkyo-ku, Tokyo 113-0033, Japan.
- d. School of Physics and Astronomy, University of St. Andrews, Fife KY16 9SS, Scotland.
- e. RIKEN Center for Emergent Matter Science, Wako, Saitama 351-0198, Japan.
- f. Materials and Structures Lab., Tokyo Institute of Technology, Yokohama, Kanagawa 226-8503, Japan
- g. Institute for Integrated Cell-Material Sciences, Kyoto University, Sakyo-ku, Kyoto 606-8501, Japan
- h. RIKEN Advanced Science Institute, Wako, Saitama 351-0198, Japan.
- i. Max-Planck-Institut für Festkörperforschung, Heisenbergstraße 1 70569 Stuttgart, Germany
- j. Institute of Advanced Industrial Science and Technology, Tsukuba, Ibaraki 305-8568, Japan.
- k. Department of Physics, Harvard University, Cambridge, MA.
- l. Dept. of Physics and Astronomy, Binghamton University, Binghamton, NY 13902.
- m. Perimeter Institute for Theoretical Physics, Waterloo, Ontario N2L 2Y5, Canada.

† These authors contributed equally to this project.

**ABSTRACT** **The identity of the fundamental broken symmetry (if any) in the underdoped cuprates is unresolved. However, evidence has been accumulating that this state may be an unconventional density wave. Here we carry out site-specific measurements within each $CuO_2$ unit-cell, segregating the results into three separate electronic structure images containing only the Cu sites ($Cu(r)$) and only the *x/y*-axis O sites ($O_x(r)$ and $O_y(r)$). Phase resolved Fourier analysis reveals directly that the modulations in the $O_x(r)$ and $O_y(r)$ sublattice images consistently exhibit a relative phase of $\pi$. We confirm this discovery on two highly distinct cuprate compounds, ruling out tunnel matrix-element and materials specific systematics. These observations demonstrate by direct sublattice phase-resolved visualization that the density wave found in underdoped cuprates consists of modulations of the intra-unit-cell states that exhibit a predominantly *d*-symmetry form factor.**

*$CuO_2$ pseudogap / broken symmetry / intra-unit-cell / density-wave form factor*

## *Electronic Inequivalence at the Oxygen Sites of the CuO$_2$ Plane in Pseudogap State*

**1**     Understanding the microscopic electronic structure of the CuO$_2$ plane represents the essential challenge of cuprate studies. As the density of doped-holes, *p*, increases from zero in this plane, the pseudogap state (1,2) first emerges, followed by the high temperature superconductivity. Within the elementary CuO$_2$ unit cell, the Cu atom resides at the symmetry point with an O atom adjacent along the *x*-axis and the *y*-axis (inset Fig. 1A). Intra-unit-cell (IUC) degrees of freedom associated with these two O sites (3,4), although often disregarded, may actually represent the key to understanding CuO$_2$ electronic structure. Among the proposals in this regard are valence-bond ordered phases having localized spin-singlets whose wavefunctions are centered on O$_x$ or O$_y$ sites (5,6), electronic nematic phases having a distinct spectrum of eigenstates at O$_x$ and O$_y$ sites (7,8), and orbital-current phases in which orbitals at O$_x$ and O$_y$ are distinguishable due to time-reversal symmetry breaking (9). A common element to these proposals is that, in the pseudogap state of lightly hole-doped cuprates, some form of electronic symmetry breaking renders the O$_x$ and O$_y$ sites of each CuO$_2$ unit-cell electronically inequivalent.

**2**     Experimental electronic structure studies that discriminate the O$_x$ from O$_y$ sites do find a rich phenomenology in underdoped cuprates. Direct oxygen site-specific visualization of electronic structure reveals that even very light hole-doping of the insulator produces local IUC symmetry breaking rendering O$_x$ and O$_y$ inequivalent (10); that both ***Q***≠0 density wave (11) and ***Q***=0 C$_4$-symmetry breaking (11,12,13) involve electronic inequivalence of the O$_x$ and O$_y$ sites; and that the ***Q***≠0 and ***Q***=0 broken symmetries weaken simultaneously with increasing *p* and disappear jointly near $p_c$=0.19 (13) For multiple cuprate compounds, neutron scattering reveals clear intra-unit-cell breaking of rotational symmetry (14,15,16). Thermal transport studies (17) can likewise be interpreted. Polarized X-ray scattering studies reveal the electronic inequivalence between O$_x$ and O$_y$ sites (18), and that angular dependent scattering is best modeled by spatially modulating their inequivalence with a *d*-symmetry form factor (19). Thus, evidence from a variety of techniques indicates that ***Q***=0 $C_4$-breaking (electronic inequivalence of $O_x$ and $O_y$) is a key element of underdoped-cuprate electronic structure. The apparently distinct phenomenology of ***Q***≠0 incommensurate density waves (DW) in underdoped cuprates has also been reported extensively (20-27). Moreover, recent studies (28,29) have demonstrated beautifully that the density modulations first visualized by STM imaging (30) are indeed the same as the DW detected by these X-ray scattering techniques. However, although distinct in terms of which symmetry is broken, there is mounting evidence that the incommensurate DW and the IUC degrees of freedom are somehow linked microscopically (13,16,19,42,44).

### *Density Waves that Modulate the CuO$_2$ Intra-unit-cell States*

***3***     One possibility is a density wave that modulates the CuO$_2$ IUC states. Proposals for such exotic DWs in underdoped cuprates include charge density waves with a *d*-symmetry form factor (31,32,33) and modulated electron-lattice coupling with a *d*-symmetry form factor (33,34). Modulations of the IUC states having wavevectors ***Q***=(*Q*,*Q*);(*Q*,-*Q*) have been extensively studied (35,36,37,38,39) but little experimental evidence for such phenomena exists. Most recently, focus has sharpened on the models (33,34,39, 40 , 41 ) yielding spatial modulations of IUC states that occur at incommensurate wavevectors ***Q***=(*Q*,0);(0,*Q*) aligned with the CuO$_2$ plane axes. The precise mathematical forms of these proposals have important distinctions, and these are discussed in full detail in SI Section I.

***4***     Density waves consisting of modulations on the $O_x$ sites that are distinct from those on the $O_y$ sites can be challenging to conceptualize. Therefore, before explaining their modulated versions, we first describe the elementary symmetry decomposition of the IUC states of CuO$_2$. There are three possibilities (SI section II): a uniform density on

*3*

the copper atoms with the $O_x$ and $O_y$ sites inactive (*s*-symmetry), a uniform density on the oxygen atoms with copper sites inactive (*s'*-symmetry), and a pattern with opposite-sign density at $O_x$ and $O_y$ sites with the copper sites inactive (*d*-symmetry). The latter state is shown in Fig. 1A. As these three IUC arrangements are spatially uniform, they correspond to specific representations of the point group symmetry of the lattice. Phase-resolved Fourier transforms of each arrangement could reveal their point group symmetry from the relative signs of the Bragg amplitudes. The *s*- and *s'*-symmetry cases both share 90º-rotational symmetry in their Bragg amplitudes (SI section II), while the Bragg amplitudes for a *d*-symmetry case change sign under 90º rotations as shown in Fig. 1B. Thus, by studying the magnitude and sign of the Bragg amplitudes in phase-resolved site specific electronic structure images, one can extract the degree to which any translationally invariant IUC arrangement has an *s*-, or *s'*- or, as in our previous work (12,13,42,43,44), a *d*-symmetry (SI Section II).

**5**     Next we consider periodic modulations of the IUC states with wavevector $\boldsymbol{Q}$, as described by

$$\rho(\boldsymbol{r}) = [S(\boldsymbol{r}) + S'(\boldsymbol{r}) + D(\boldsymbol{r})]Cos(\boldsymbol{Q} \cdot \boldsymbol{r} + \phi(\boldsymbol{r})) \quad (1)$$

Here $\rho(\boldsymbol{r})$ is a generalized density representing whatever electronic degree of freedom is being modulated; *S, S'* and *D* are the coefficients of the DW form factors with *s*-, *s'*- and *d*-symmetry respectively, and $\phi(\boldsymbol{r})$ is an overall phase (that might be spatially disordered (42,44)). A simple way to understand these DW form factors is in terms of the three $CuO_2$ sublattices: $\boldsymbol{r}_{Cu}$, $\boldsymbol{r}_{O_x}$, $\boldsymbol{r}_{O_y}$ (inset Fig. 1A). By definition, $S(\boldsymbol{r}) = A_S$ if $\boldsymbol{r} \in \boldsymbol{r}_{Cu}$ and otherwise zero; $S'(\boldsymbol{r}) = A'_S$ if $\boldsymbol{r} \in \boldsymbol{r}_{O_x}$ or $\boldsymbol{r}_{O_y}$ and otherwise zero; $D(\boldsymbol{r}) = A_D$ if $\boldsymbol{r} \in \boldsymbol{r}_{O_x}$; $D(\boldsymbol{r}) = -A_D$ if $\boldsymbol{r} \in \boldsymbol{r}_{O_y}$ and otherwise zero. This last case is a *d*-form factor density wave (*d*FF-DW) as shown schematically in Fig. 1C. In cuprates, a generic DW can actually have *S, S'*, and *D* all non-zero because the directionality of modulation wavevector $\boldsymbol{Q}$ breaks rotational symmetry (SI Section II).   Therefore, to identify a predominantly *d*FF-DW one should consider the structure of its Fourier transform, which exhibits several distinctive features. Figure 1C shows a schematic *d*FF-DW that modulates along $\boldsymbol{Q}_x$. In this state, by considering the two trajectories parallel to $\boldsymbol{Q}_x$ marked O$_x$ and O$_y$, we see that amplitude of the wave along O$_x$ is exactly π out of phase with that along the adjacent trajectory O$_y$. For this reason, when its Fourier transform is determined (Fig. 1D), no primary modulation peaks occur at ± $\boldsymbol{Q}_x$ inside the first Brillouin zone (BZ). The second effect is that the Bragg satellite peaks at $\boldsymbol{Q}' = (1,0)\pm\boldsymbol{Q}_x$ and $\boldsymbol{Q}''=(0,1)\pm\boldsymbol{Q}_x$ have opposite sign as shown in Fig. 1D (SI Section III). Note, that if an equivalent *d*FF-DW occurred only along $\boldsymbol{Q}_y$ (⊥ $\boldsymbol{Q}_x$), the effects would correspond to those in Fig. 1D but the Bragg satellite peaks then occurring at $\boldsymbol{Q}' = (1,0)\pm\boldsymbol{Q}_y$ and $\boldsymbol{Q}''=(0,1)\pm\boldsymbol{Q}_y$ and again having opposite sign (SI Section III).

## *Sublattice Phase-Resolved Fourier Transform STM*

*6*      With the recent development of STM techniques to image IUC electronic structure (10,11,12,13,44) while simultaneously achieving high-precision phase-resolved Fourier analysis (12,13,42,44), it was suggested by one of us (S.S.) that a practical approach to determining the form factor of underdoped cuprate DWs would be to partition each such image of the $CuO_2$ electronic structure, into three separate images. The first image contains only the measured values at Cu sites ($Cu(\boldsymbol{r})$) and the other two images $O_x(\boldsymbol{r})$ and $O_y(\boldsymbol{r})$ only the measurements at the x/y-axis oxygen sites. The latter are key because two types of DW: $\rho_{S'}(\boldsymbol{r}) = S'Cos(\boldsymbol{Q}\cdot\boldsymbol{r}+\phi_{S'}(\boldsymbol{r}))$ and $\rho_D(\boldsymbol{r}) = DCos(\boldsymbol{Q}\cdot\boldsymbol{r}+\phi_D(\boldsymbol{r}))$ are both actually formed by using only phenomena from the $O_x/O_y$ sites (Fig. 1C). Once the original electronic structure image is thus separated, the phase-resolved Fourier transform of $O_x(\boldsymbol{r})$ and $O_y(\boldsymbol{r})$, $\tilde{O}_x(\boldsymbol{q})$ and $\tilde{O}_y(\boldsymbol{q})$, may, in principle, be used to reveal the form factor of these two types of DW.

*7*      A *d*FF-DW with modulations along both x- and y-axes at $\boldsymbol{Q}=(Q,0);(0,Q)$ should then exhibit two key characteristics exemplified by Fig. 1E whose equivalent experimental information is contained in $Re\tilde{O}_x(\boldsymbol{q}) + Re\tilde{O}_y(\boldsymbol{q})$ (SI Section III). The first is that modulation peaks at $\boldsymbol{Q}$ should disappear in $Re\tilde{O}_x(\boldsymbol{q}) + Re\tilde{O}_y(\boldsymbol{q})$ while the Bragg-satellite peaks at $\boldsymbol{Q'} = (1,0)\pm\boldsymbol{Q}$ and at $\boldsymbol{Q''}=(0,1)\pm\boldsymbol{Q}$ should exist with opposite sign as shown in Fig. 1E (the same being true for $Im\tilde{O}_x(\boldsymbol{q}) + Im\tilde{O}_y(\boldsymbol{q})$). The second predicted characteristic is that the primary DW peaks at $\boldsymbol{Q}$ should exist clearly in $Re\tilde{O}_x(\boldsymbol{q}) - Re\tilde{O}_y(\boldsymbol{q})$ while their Bragg-satellite peaks at $\boldsymbol{Q'} = (1,0)\pm\boldsymbol{Q}$ and $\boldsymbol{Q''}=(0,1)\pm\boldsymbol{Q}$ should disappear (the same being true for $Im\tilde{O}_x(\boldsymbol{q}) - Im\tilde{O}_y(\boldsymbol{q})$.) This is required because, if all $O_y$ sites are multiplied by -1 as when we take the difference $Re\tilde{O}_x(\boldsymbol{q}) - Re\tilde{O}_y(\boldsymbol{q})$, a $d$FF-DW is converted to a $s'$-form factor DW. For this reason, the signature of a $d$FF-DW in $Re\tilde{O}_x(\boldsymbol{q}) - Re\tilde{O}_y(\boldsymbol{q})$ is that it should exhibit the characteristics of Fig. 1F (SI Section III).

*Experimental Methods*

**8** To search for such phenomena, we use SI-STM (44) to measure both the differential tunneling conductance $\frac{dI}{dV}(\boldsymbol{r}, E = eV) \equiv g(\boldsymbol{r}, E)$ and the tunnel-current magnitude $I(\boldsymbol{r}, E)$, at bias voltage $V$, and on samples of both Bi$_2$Sr$_2$CaCu$_2$O$_{8+x}$ (BSCCO) and Ca$_{2-x}$Na$_x$CuO$_2$Cl$_2$ (NaCCOC). Because the electronic density-of-states $N(\boldsymbol{r}, E)$ always enters as $g(\boldsymbol{r}, E) \propto \left[eI_s / \int_0^{eV_s} N(\boldsymbol{r}, E')dE'\right] N(\boldsymbol{r}, E)$ where $I_s$ and $V_s$ are arbitrary parameters, the unknown denominator $\int_0^{eV_s} N(\boldsymbol{r}, E')dE'$ always prevents valid determination of $N(\boldsymbol{r}, E)$ based only upon $g(\boldsymbol{r}, E)$ measurements. Instead, $Z(\boldsymbol{r}, |E|) = g(\boldsymbol{r}, E)/g(\boldsymbol{r}, -E)$ or $R(\boldsymbol{r}, |E|) = I(\boldsymbol{r}, E)/-I(\boldsymbol{r}, -E)$ are used (11,12,13,42,44) in order to suppress the otherwise profound systematic errors. This approach allows distances, wavelengths, and phases of electronic structure to be measured correctly. Physically, the ratio $R(\boldsymbol{r}, V) \propto \int_o^{eV} N(\boldsymbol{r}, E)dE / \int_{-eV}^{0} N(\boldsymbol{r}, E)dE$ is measured using an identical tip-sample tunnel junction formed at $\boldsymbol{r}$ but using opposite bias voltage $\pm V$; it is a robust measure of the spatial symmetry of electronic states in the energy range $|E|=eV$. Additionally for this study, measurements at many pixels within each UC are required (to spatially discriminate every $O_x$, $O_y$ and Cu site) while simultaneously measuring in a sufficiently large FOV to achieve high resolution in phase definition (11,12,43,44).

**9** Data acquired under these circumstance are shown in Figure 2A, the measured $R(r, |E|=150\text{meV})$ for a BSCCO sample with $p\sim 8\pm 1\%$. This FOV contains $\sim$ 15,000 each of individually resolved Cu, $O_x$ and $O_y$ sites. Figure 2B shows a magnified part of this $R(r)$ with Cu sites indicated by blue dots; Figure 2C is the simultaneous topographic image showing how to identify the coordinate of each Cu, $O_x$ and $O_y$ site in all the images. Using the Lawler-Fujita phase-definition algorithm which was developed for IUC symmetry determination studies (12,43,44) we achieve a phase accuracy of $\sim 0.01\pi$ (43) throughout. Figure 2D,E show the partition of measured $R(r)$ into two oxygen-site-specific images $O_x(r)$ and $O_y(r)$ determined from Fig. 2B (segregated Cu-site specific image is shown SI Section V). Larger FOV $O_x(r);O_y(r)$ images partitioned from $R(r)$ in Fig. 2A, and their Fourier transforms are shown in SI Section V.

***Direct Measurement of the DW Form Factor from Sublattice Phase-Resolved Images***

**10** Now we consider the complex Fourier transforms of $O_x(r)$ and $O_y(r)$, $\tilde{O}_x(q)$ and $\tilde{O}_y(q)$, as shown in Fig. 3A,B. We note that the use of $R(r,V)$ or $Z(r,V)$ is critically important for measuring relative phase of $O_x/O_y$ sites throughout any DW, because analysis of $g(r,V)$ shows how the tip-sample junction establishment procedure (11,44) scrambles the phase information irretrievably. Upon calculating the sum $Re\tilde{O}_x(q) + Re\tilde{O}_y(q)$ as shown in Fig. 3C, we find no DW modulation peaks in the vicinity of $Q$. Moreover there is evidence for a $\pi$-phase shift between much sharper peaks at $Q'$ and $Q''$ (albeit with phase disorder). Both of these effects are exactly as expected for a $d$FF-DW (see Fig. 1E,F). Further, the modulation peak at $Q$ inside the first BZ that is weak in Figs. 3A and 3B and absent in Fig. 3C is clearly visible in $Re\tilde{O}_x(q) - Re\tilde{O}_y(q)$ as shown in Fig. 3D. Hence the absence of this feature in $Re\tilde{O}_x(q) + Re\tilde{O}_y(q)$ cannot be ascribed to broadness of the features surrounding $q = 0$; rather, it is due to a high fidelity phase cancelation between the modulations on $O_x$ and $O_y$ occurring with $q\sim Q$. Finally, the Bragg-satellite peaks at $Q' = (1,0)\pm Q$ and $Q''=(0,1)\pm Q$ that were clear in $Re\tilde{O}_x(q) + Re\tilde{O}_y(q)$ are absent in $Re\tilde{O}_x(q) - Re\tilde{O}_y(q)$. Comparison of all these observations with predictions for a $d$FF-DW in Fig. 1E,F demonstrates that the modulations at $Q$ maintain a phase difference of approximately $\pi$ between $O_x$ and $O_y$ within each unit cell, and therefore predominantly constitute a $d$-form factor DW.

**11** To demonstrate that these phenomena are not a specific property of a given tip-sample tunnel matrix element, or crystal symmetry, or surface termination layer, or cuprate material family, we carry out the identical analysis on data from NaCCOC samples with $p\sim 12\pm 1\%$ (SI Section V). For this compound, Fig. 3E,F are the measured $Re\tilde{O}_x(q)$ and $Re\tilde{O}_y(q)$. Again, the absence of DW peaks at $Q$ in Fig. 3G which shows $Re\tilde{O}_x(q) + Re\tilde{O}_y(q)$ are due to cancelation between $O_x$ and $O_y$ contributions, as these peaks are visible in $Re\tilde{O}_x(q)$ and $Re\tilde{O}_y(q)$ (Figs. 3E,F). Moreover, the sign change between the Bragg satellites $Q' = (1,0)\pm Q$ and $Q''=(0,1)\pm Q$ in Fig. 3G is another hallmark of a $d$FF-DW. Finally, $Re\tilde{O}_x(q) - Re\tilde{O}_y(q)$ reveals again that the modulation peaks at $Q$ inside the first BZ that are invisible in Fig. 3G become vivid in Fig. 3H, while the Bragg-satellites disappear. One can see directly that these results are comprehensively consistent with observations in Figs 3A-D meaning that the DW of NaCCOC also exhibits a robustly $d$-symmetry form factor. This observation rules out experimental/materials systematics as the source of the $d$FF-DW signal and therefore signifies that this state is a fundamental property of the underdoped $CuO_2$ plane.

### Cuprate d-Form Factor DW is both Predominant and Robust

*12*     One can quantify the preponderance of the $d$FF-DW by measuring the magnitude of the $s$-, $s'$- and $d$-symmetry form factors of the observed DW near $\boldsymbol{Q}$ inside the first BZ (SI Section II). In Fig. 4A we show the power spectral density Fourier transform analysis $|(\tilde{O}_x(\boldsymbol{q}) - \tilde{O}_y(\boldsymbol{q}))/2|^2$ yielding the $d$-form factor magnitude $D$, with the equivalent results for only the Cu sites $|\tilde{C}_u(\boldsymbol{q})|^2$ to determine the $s$-form factor magnitude $S$, and $|(\tilde{O}_x(\boldsymbol{q}) + \tilde{O}_y(\boldsymbol{q}))/2|^2$ for the $s'$-form factor magnitude $S'$, shown in SI Section V. In Fig. 4B, the measured values $S$, $S'$ and $D$ are plotted along the dashed lines through $\boldsymbol{Q}$ in Fig. 4A, and shows that the $d$-form factor component is far stronger than the others for both modulation directions. This is also the case in the NaCCOC data. Figure 4C shows examples of measured complex valued $\tilde{O}_x(\boldsymbol{q}) \equiv Re\tilde{O}_x(\boldsymbol{q}) + iIm\tilde{O}_x(\boldsymbol{q})$ and compares them to $\tilde{O}_y(\boldsymbol{q}) \equiv Re\tilde{O}_y(\boldsymbol{q}) + iIm\tilde{O}_y(\boldsymbol{q})$ for each of a series of representative $\boldsymbol{q}$ within the DW peaks surrounding $\boldsymbol{Q}$ (all such data are from Figs 2,3). Figure 4D is a 2D-histogram showing both the magnitude difference and the phase difference between all such pairs of Fourier-filtered $\tilde{O}_x(\boldsymbol{r}):\tilde{O}_y(\boldsymbol{r})$ within the same broad DW peaks (SI Section V). These data reveal the remarkably robust nature of the $d$-form factor of the DW, and that the strong spatial disorder in DW modulations (e.g. Fig. 2A and Ref. 42) has little impact on the phase difference of $\pi$ between $O_x$ and $O_y$ within every $CuO_2$ unit cell. Finally, focusing on specific regions of the $R(\boldsymbol{r})$ images, one can now understand in microscopic detail the well-known (11,13,42,44) but unexplained spatial patterns of $CuO_2$ electronic structure when detected with sub-unit-cell resolution (e.g. Fig. 4E). In fact, the virtually identical electronic structure patterns in BSCCO and NaCCOC (Fig. 4E, Ref. 11) correspond to the instance in which a unidirectional $d$FF-DW occurs locally with $\boldsymbol{Q}$=(0.25,0) and with amplitude peaked on the central $O_x$ sites (dashed vertical arrow).  A model of a $d$FF-DW with this choice of spatial-phase is shown in Fig. 4F (SI Section I) with the calculated density adjacent; the agreement between data (Fig. 4E) and *d*FF-DW model (Fig. 4F) gives a clear visual confirmation that the patterns observed in real space $R(\mathbf{r})$ data (10,11,13,44) are a direct consequence of a locally commensurate, unidirectional, *d*-form factor density wave.

## *Conclusions and Discussion*

**13**    By generalizing our technique of phase-resolved intra-unit-cell electronic structure imaging at Cu, $O_x$, $O_y$ (11,12,13,42,43,44), to include segregation of such data into three images ($Cu(\mathbf{r})$, $O_x(\mathbf{r})$, $O_y(\mathbf{r})$), sublattice-phase-resolved Fourier analysis yielding $\widetilde{Cu}(\mathbf{q})$, $\tilde{O}_x(\mathbf{q})$ and $\tilde{O}_y(\mathbf{q})$ becomes possible. Then, by comparing predicted signatures of a *d*FF-DW in Figs. 1E,F with the equivalent measurements $Re\ \tilde{O}_x(\mathbf{q}) \pm Re\ \tilde{O}_y(\mathbf{q})$ in Fig. 3C,D and Fig. 3G,H respectively, we find them in excellent agreement for both BSCCO and NaCCOC. In their X-ray observations on underdoped $Bi_2Sr_2CuO_6$ and $YBa_2Cu_3O_7$, Comin *et al* (19) used a model of the scattering amplitudes of the Cu and O atoms in the presence of charge-density modulations, and showed that a density wave that modulates with a *d*-form factor between $O_x$ and $O_y$ sites, provides a significantly better fit to the measured cross section than *s*- or *s'*-form factors. In our complementary approach, we demonstrate using direct sublattice phase-resolved visualization that the cuprate density waves involve modulations of electronic structure (not necessarily charge density) that maintain a relative phase of ~$\pi$ between $O_x$ and $O_y$, or equivalently that exhibit a *d*-symmetry form factor.

**14**    New challenges are revealed by the detection of the dFF-DW state in the cuprate pseudogap phase. The first is to determine its relationship to the $\mathbf{Q}$=0 $C_4$-breaking phenomena that are widely reported in underdoped cuprates (12-18). In this regard, a potentially significant observation is that, coupling between a DW with both s'- and *d*-form factor components and a ***Q***=0 nematic state is allowed by symmetry at the level of Ginzburg-Landau theory, so that their coexistence is not unreasonable. Nevertheless, the specifics of that coupling (42), and how the phase diagram is arranged (13) in terms of the *d*FF-DW and ***Q***=0 states remains to be determined and understood. Another challenge is to understand the microscopic physics of the *d*FF-DW itself. The fidelity of the $\pi$ phase difference within each unit-cell (Fig. 4) despite the intense spatial disorder (Fig. 2) implies that there must be a powerful microscopic reason for inequivalence of electronic structure at the $O_x$ and $O_y$ sites in underdoped cuprates. One may ask whether, in addition to antiferromagnetic interactions (34-41), the Coulomb interactions between holes on adjacent $O_x$ and $O_y$ sites that play a key role in theories for the ***Q***=0 nematic (7,8,38), may be an important factor in formation of the *d*FF-DW. Finally, finding the microscopic relationship of the *d*FF-DW to the *d*-wave high temperature superconductivity and to the pseudogap state now emerge as issues of primary importance.

**Figure Captions**

### Figure 1 Intra-unit-cell Electronic Structure Symmetry in the CuO$_2$ Plane

A. Schematic of translationally invariant electronic structure pattern with opposite-sign density on O$_x$ and O$_y$ (*d*-symmetry), as discussed in Refs 12,13,44. The inactive Cu sites are indicated by grey dots. Inset: Elementary C$_u$, O$_x$ and O$_y$ orbitals, on the three sublattices in the CuO$_2$ plane; $r_{Cu}$ sublattice contains only the Cu sites, $r_{Ox}$ sublattice contains only the O$_x$ sites and $r_{Oy}$ sublattice contains only the O$_y$ sites. Referring to this arrangement as having '*d*-symmetry' does not, at this stage, imply any specific relationship to the d-symmetry Cooper pairing of the superconducting state.

B. Fourier transform of the Q=0 $C_4$-breaking pattern of opposite-sign density on O$_x$ and O$_y$ in A. The Bragg peaks have the opposite sign indicating the IUC states have *d*-symmetry (12,44).

C. Schematic of a *d*-form factor DW $\rho_D(r) = D Cos(Q_x \cdot r + \phi_D(r))$ with the origin chosen at a Cu site and modulating only along the *x*-axis. The colors represent the density $\rho_D(r)$ at every oxygen site. Two fine lines, one labeled O$_x$ traverses parallel to $Q_x$ and passing only through oxygen sites oriented along the *x*-axis, and the second one labeled O$_y$ again traverses parallel to $Q_x$ but passing only through oxygen sites oriented along the y-axis, are shown. The color scale demonstrates how the amplitude of the DW is exactly π out of phase along these two trajectories.

D. The real-component of the Fourier transform of the pattern in (1C). For this *d*-form factor DW, the Bragg-satellite peaks at ***Q'*** and ***Q"*** exhibit opposite sign. More profoundly, because they are out of phase by π, the modulations of O$_x$ and O$_y$ sites cancel, resulting in the disappearance of the DW modulation peaks at ±***Q*** within the BZ (dashed box).

*12*

E. The real-component of the Fourier transform of a *d*-form factor DW having modulations at $Q=(Q,0);(0,Q)$ (SI Section III); the DW satellites of inequivalent Bragg peaks at $Q'$ and $Q''$ exhibit opposite sign, and the basic DW modulation peaks at $Q$ have disappeared from within the BZ, as indicated by empty circles.

F. The real-component of the Fourier transform of an *s'*-form factor DW having modulations at $Q=(Q,0);(0,Q)$ (SI Section III) ; the DW satellites of inequivalent Bragg peaks at $Q'$ and $Q''$ now cancel while the basic DW modulation peaks at $Q$ are robust within the BZ.

**Figure 2 Oxygen-site-specific Imaging and Segregation of R(*r*)**

A. Measured $R(r)$ with ~16 pixels within each $CuO_2$ unit cell and ~45 nm square FOV for BSCCO sample with $p$~8±1%. This R(*r*) electronic structure image reveals the extensive local IUC $C_4$-breaking (12,13) (SI Section V).

B. Smaller section of R(*r*) in FOV of 2A, now showing the location of the Cu lattice as blue dots. The well-known (11,12,13,44) breaking of $C_4$ rotational symmetry within virtually every $CuO_2$ unit cell and the modulations thereof, are obvious.

C. Topographic image of FOV in 2B showing Cu lattice sites as identified from the Bi atom locations as blue dots. By using the Lawler-Fujita algorithm (12,43) spatial-phase accuracy for the $CuO_2$ plane of ~0.01π is achieved throughout.

D. In the same FOV as 2B, we measure the value of *R* at every $O_x$ site and show the resulting function $O_x(r)$.

E. In the same FOV as 2B, we measure the value of *R* at every $O_y$ site and show the resulting function $O_y(r)$.

**Figure 3 Sublattice Phase-resolved Analysis Revealing *d*-Form Factor DW**

A. Measured $Re\tilde{O}_x(\boldsymbol{q})$ from R(***r***) in 2A; the four DW peaks at ***Q***, and the DW Bragg-satellite peaks exist but are all poorly resolved.

B. $Re\tilde{O}_y(\boldsymbol{q})$ from 2A; again, the four DW peaks at ***Q***, and the DW Bragg-satellite peaks exist but are all poorly resolved.

C. Measured $Re\tilde{O}_x(\boldsymbol{q}) + Re\tilde{O}_y(\boldsymbol{q})$ from A,B. The four DW peaks at ***Q*** are not detectable while the DW Bragg-satellite peaks are enhanced and well resolved. Comparing to Fig. 1E these are the expected phenomena of a *d*-form factor DW (with spatial disorder in $\phi_D$).

D. Measured $Re\tilde{O}_x(\boldsymbol{q}) - Re\tilde{O}_y(\boldsymbol{q})$ from A,B. The primary DW peaks at ***Q*** are strongly enhanced while the DW Bragg-satellite peaks have disappeared. Comparing to Fig. 1F, these are once again the expected phenomena of a *d*FF-DW.

E. Measured $Re\tilde{O}_x(\boldsymbol{q})$ for NaCCOC sample with *p*~12±1%; the DW peaks at ***Q***, and the DW Bragg-satellite peaks exist but are poorly resolved.

F. Measured $Re\tilde{O}_y(\boldsymbol{q})$ for NaCCOC; the DW peaks at ***Q***, and the DW Bragg-satellite peaks exist but are poorly resolved.

G. Measured $Re\tilde{O}_x(\boldsymbol{q}) + Re\tilde{O}_y(\boldsymbol{q})$ from E,F. The four DW peaks at ***Q*** are no longer detectable while the DW Bragg-satellite peaks are enhanced and well resolved. Importantly (modulo some phase noise) the Bragg-satellite peaks at inequivalent ***Q'*** and ***Q"*** exhibit opposite sign. Comparing to Fig. 1E these are the expected phenomena of a *d*FF-DW.

H. Measured $Re\tilde{O}_x(\boldsymbol{q}) - Re\tilde{O}_y(\boldsymbol{q})$ from E, F. The four DW peaks at ***Q*** are enhanced while the DW Bragg-satellite peaks have disappeared. Comparing to Fig. 1F these confirm the *d*FF-DW conclusion.

**Figure 4 *d*-Form Factor DW: Predominance and Robustness**

A. PSD Fourier transforms of $R(r)$ measured only at the $O_x/O_y$ sites yielding $|(\tilde{O}_x(q) - \tilde{O}_y(q))/2|^2$. This provides the measure of relative strength of the *d*-form factor in the DW.

B. Measured PSD is plotted along the dashed line through $Q$ in Fig. 4A and shows the *d*-form factor component predominates greatly. The measured ratios within the DW peaks surrounding $Q$ is $D/S > 5$ and $D/S' > 12$.

C. $\tilde{O}_x(q) \equiv Re\tilde{O}_x(q) + iIm\tilde{O}_x(q)$ compared to $\tilde{O}_y(q) \equiv Re\tilde{O}_y(q) + iIm\tilde{O}_y(q)$ for each of a series of representative $q$ within the DW peaks surrounding $Q$. This shows how, wherever the $CuO_2$ unit cell resides in the disordered DW (Fig 2A), the relative phase between the $O_x$ and $O_y$ sites is very close to $\pi$ while the difference in magnitudes are close to zero.

D. Two-axis histogram of difference in normalized magnitude (vertical) and phase (horizontal) between all pairs $\tilde{O}_x(r, Q)$ and $\tilde{O}_y(r, Q)$ which are obtained by Fourier filtration of $O_x(r)$ and $O_y(r)$ to retain only $q \sim Q$ (SI Section V). This represents the measured distribution of amplitude difference, and phase difference, between each pair of $O_x/O_y$ sites everywhere in the DW. It demonstrates directly that their relative phase is always close to $\pi$ and that their magnitude differences are always close to zero.

E. Measured $R(r)$ images of local electronic structure patterns that commonly occur in BSCCO and NaCCOC (11). The Cu and $O_x$ sites (as labeled by solid and dashed arrows respectively) were determined independently and directly from topographic images (11).

F. *d*FF-DW model with $Q=(0.25,0)$ and amplitude maximum on the central $O_x$ site (dashed arrow); the calculated charge density pattern from this model is shown adjacent. Therefore a *d*FF-DW model with this particular spatial-phase provides an excellent explanation for the observed density patterns shown in E and reported previously in Refs11,12,13 and 44.

**Acknowledgements**

We acknowledge and thank S. Billinge, R. Comin, A. Damascelli, D.-H. Lee, S.A. Kivelson, A. Kostin, and A.P. Mackenzie, for very helpful discussions and communications. Experimental studies were supported by the Center for Emergent Superconductivity, an Energy Frontier Research Center, headquartered at Brookhaven National Laboratory and funded by the U.S. Department of Energy under DE-2009-BNL-PM015, as well as by a Grant-in-Aid for Scientific Research from the Ministry of Science and Education (Japan) and the Global Centers of Excellence Program for Japan Society for the Promotion of Science. C. K. K. acknowledges support from the *Fluc Team* program at Brookhaven National Laboratory under contract DE-AC02-98CH10886. S.D.E. acknowledges the support of EPSRC through the Programme Grant "Topological Protection and Non-Equilibrium States in Correlated Electron Systems". Y.K. acknowledges support form studies at RIKEN by JSPS KAKENHI (19840052, 20244060). Theoretical studies at Cornell University were supported by NSF Grant DMR-1120296 to the Cornell Center for Materials Research and by NSF Grant DMR-0955822. A.A. and S.S. are supported by NSF Grant DMR-1103860 and by the Templeton Foundation.

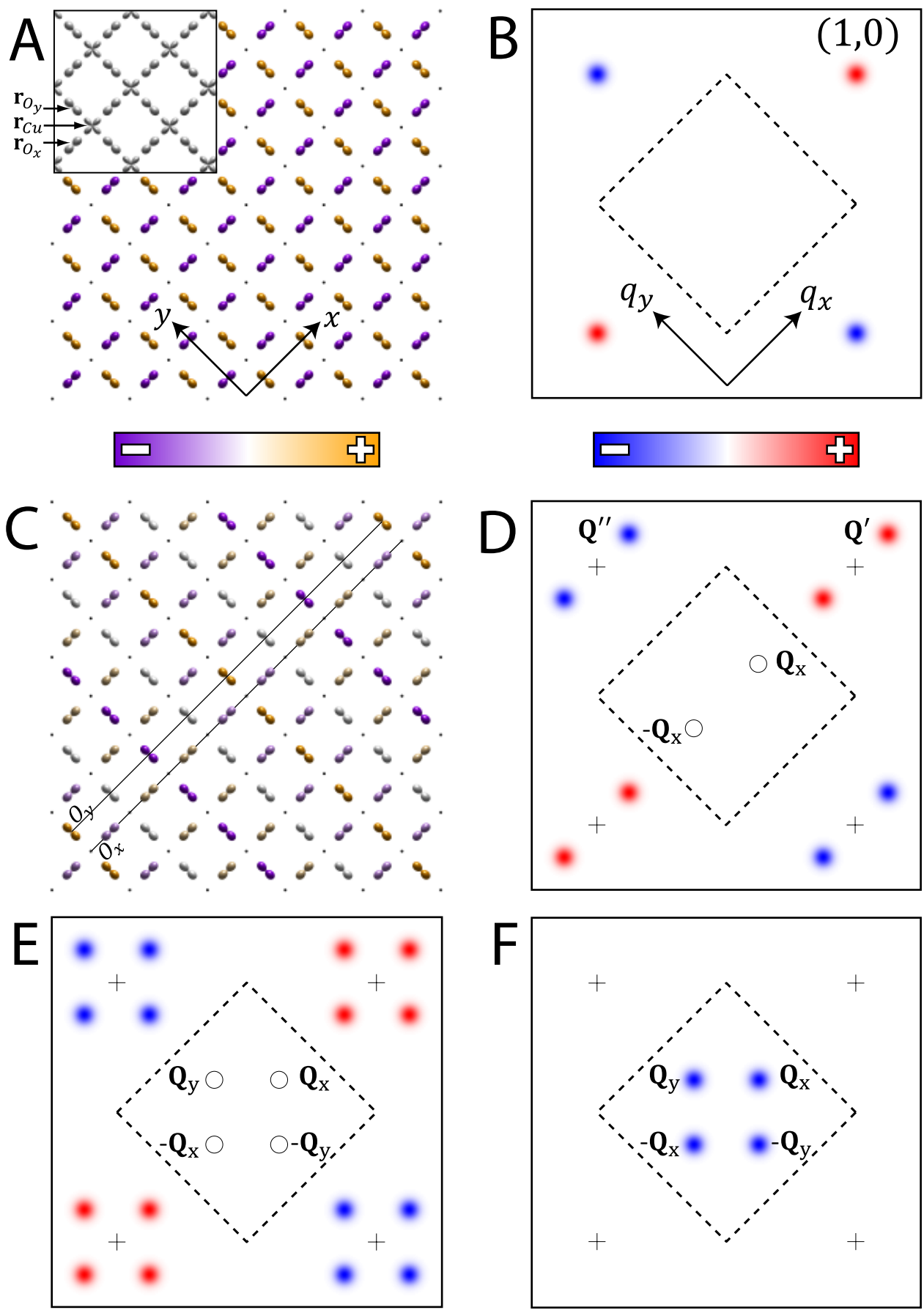

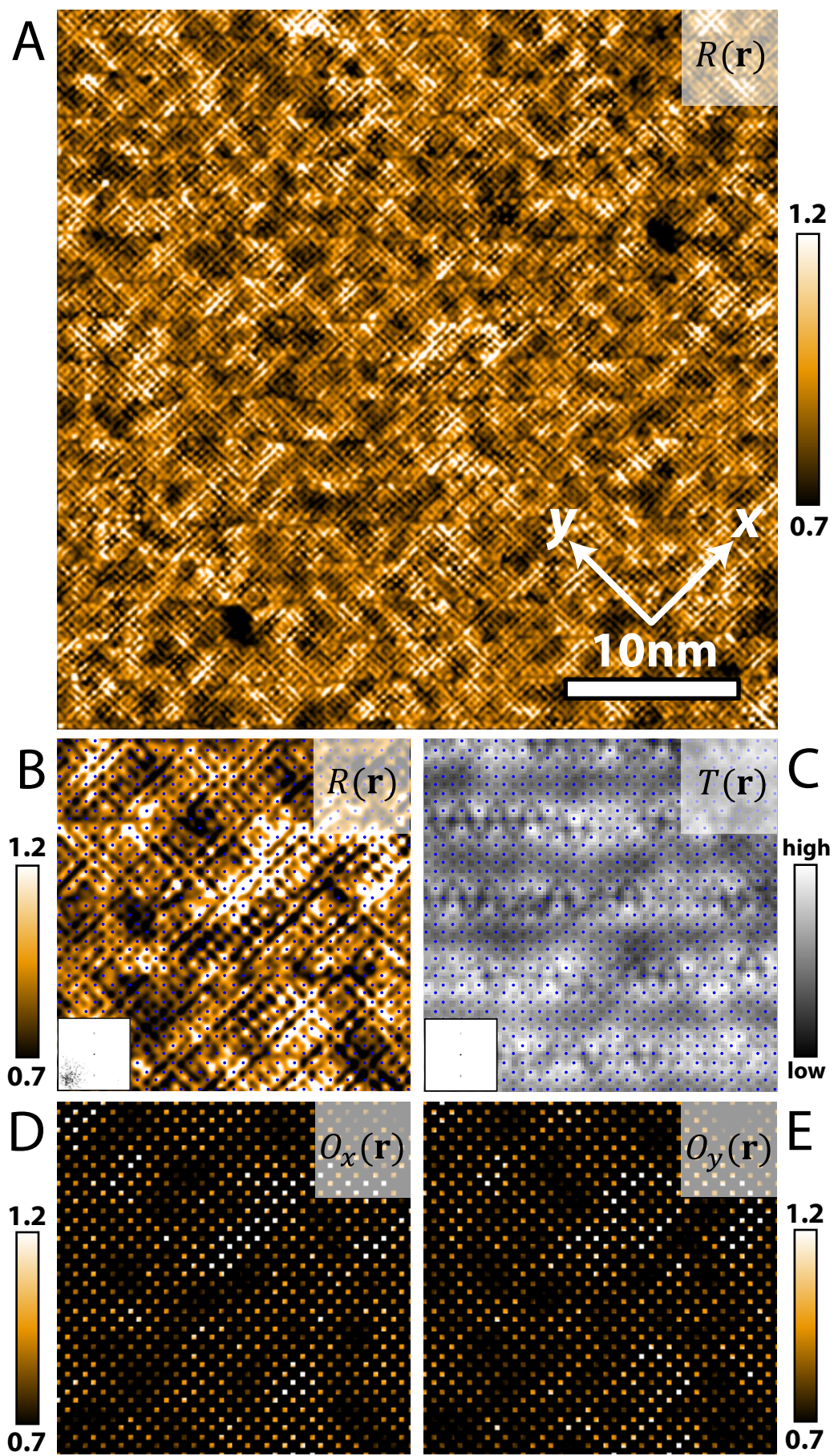

FIG2

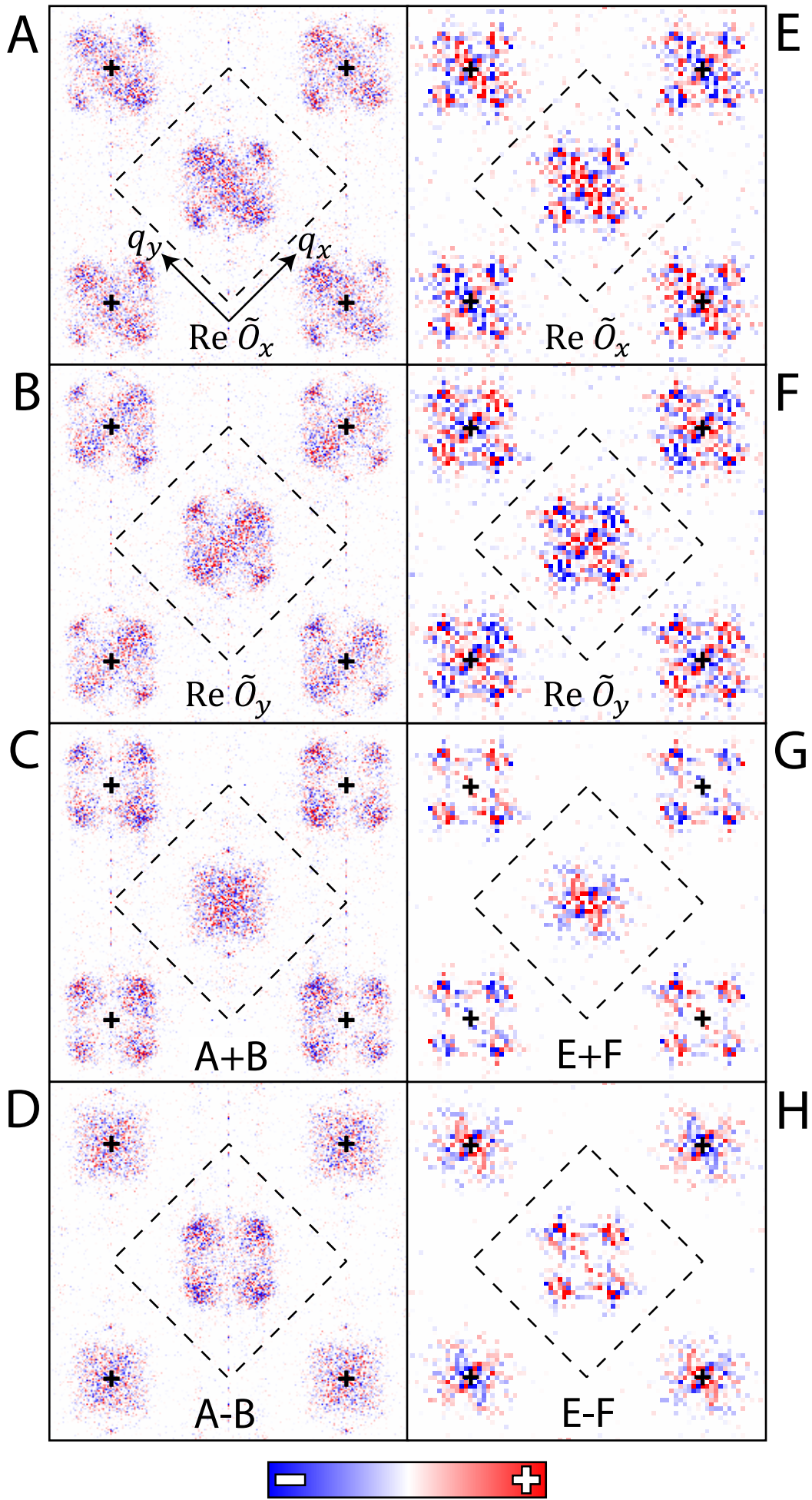

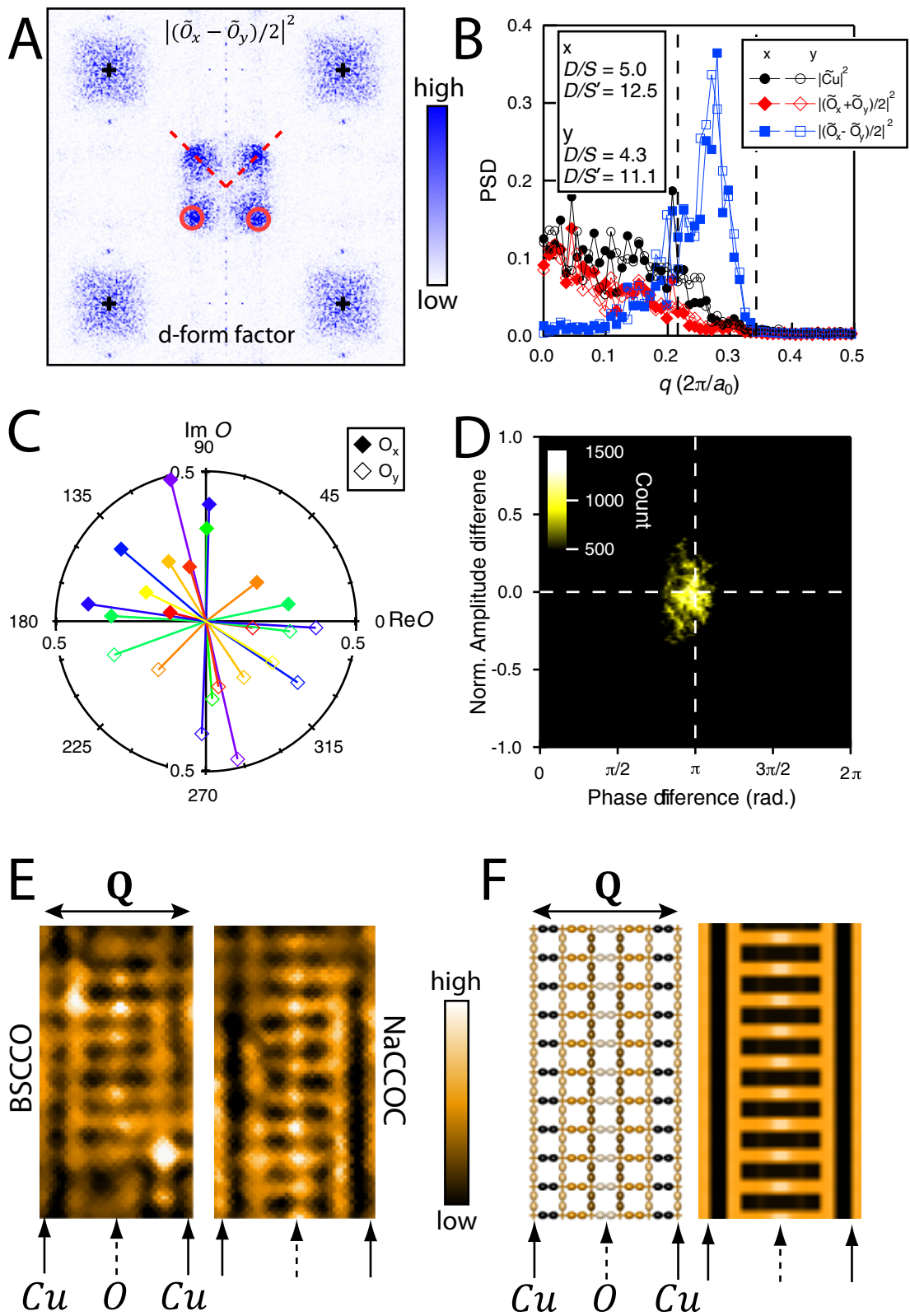

*Supporting Information For*

# Direct phase-sensitive identification of a *d*-form factor density wave in underdoped cuprates

K. Fujita†, M. H. Hamidian†, S. D. Edkins, Chung Koo Kim, Y. Kohsaka,

M. Azuma, M. Takano, H. Takagi, H. Eisaki, S. Uchida, A. Allais,

M. J. Lawler, E. -A. Kim, Subir Sachdev, & J. C. Séamus Davis

## I  Models of spatially modulated order in underdoped cuprates

The study of the underdoped cuprates has led to proposals of a large number of density-wave ordered states with non-trivial form factors [1]-[33]. Here we provide a unified perspective on these orders, while highlighting the key characteristics detected by our observations.

It is useful to begin by considering the following order parameter at the Cu sites $\mathbf{r}_i$ and $\mathbf{r}_j$ [21, 22]

$$\langle c^\dagger_{i\alpha} c_{j\alpha} \rangle = \sum_Q \left[ \sum_k P(\mathbf{k}, \mathbf{Q}) e^{i\mathbf{k}\cdot(\mathbf{r}_i - \mathbf{r}_j)} \right] e^{i\mathbf{Q}\cdot(\mathbf{r}_i + \mathbf{r}_j)/2} \qquad (1.1)$$

where $c_{i\alpha}$ annihilates an electron with spin $\alpha$ on a site at position $\mathbf{r}_i$. Here the wavevector $\mathbf{Q}$ is associated with a modulation in the *average* co-ordinate $(\mathbf{r}_i + \mathbf{r}_j)/2$. The form-factor describes the dependence on the *relative* co-ordinate $\mathbf{r}_i - \mathbf{r}_j$. An advantage of the formulation in Eq. (1.1) is that it provides an efficient characterization of symmetries. The operator identity $\langle A^\dagger \rangle \equiv \langle A \rangle^*$ requires that

$$P^*(\mathbf{k}, \mathbf{Q}) = P(\mathbf{k}, -\mathbf{Q}) \qquad (1.2)$$

while

$$P(\mathbf{k}, \mathbf{Q}) = P(-\mathbf{k}, \mathbf{Q}) \qquad (1.3)$$

if time-reversal symmetry is preserved.

A number of other studies [1, 4, 16, 17, 19] have made the closely related, but distinct parameterization

$$\langle c^\dagger_{i\alpha} c_{j\alpha} \rangle = \sum_Q \left[ \sum_k f(\mathbf{k}, \mathbf{Q}) e^{i\mathbf{k}\cdot(\mathbf{r}_i - \mathbf{r}_j)} \right] e^{i\mathbf{Q}\cdot\mathbf{r}_i} \qquad (1.4)$$

and then considered various ansatzes for the function $f(\mathbf{k}, \mathbf{Q})$. These are clearly related to those for

$P(\mathbf{k}, \mathbf{Q})$ by

$$f(\mathbf{k}, \mathbf{Q}) = P(\mathbf{k} + \mathbf{Q}/2, \mathbf{Q}) \qquad (1.5)$$

It is now clear that the relations (1.2) and (1.3) take a more complex form in terms of $f(\mathbf{k}, \mathbf{Q})$. Also, a $d$-wave form factor for $f(\mathbf{k}, \mathbf{Q})$ is not equal to a $d$-wave form factor for $P(\mathbf{k}, \mathbf{Q})$, except at $\mathbf{Q} = 0$.

We conduct the remainder of the discussion using $P(\mathbf{k}, \mathbf{Q})$ and Eq. (1.1). Depending upon the value of $\mathbf{Q}$, various crystalline symmetries can also place restrictions on $P(\mathbf{k}, \mathbf{Q})$, and we illustrate this with a few examples.

An early discussion of a state with non-trivial form factors was the "staggered flux" state (also called the "$d$-density wave" state), which carries spontaneous staggered currents [2]-[6]. This state has $P(\mathbf{k}, \mathbf{Q})$ non-zero only for $\mathbf{Q} = (\pi, \pi)$, and

$$P(\mathbf{k}, \mathbf{Q}) = P_{sf}(\sin(k_x) - \sin(k_y)) + P'_{sf}(\sin(2k_x) - \sin(2k_y)) + \cdots \qquad (1.6)$$

where $P_{sf}$, $P'_{sf}$ are constants. All terms on the right-hand-side are required by symmetry to be odd under time-reversal (*i.e.* odd in $\mathbf{k}$), and odd under the interchange $k_x \leftrightarrow k_y$. In the present notation, therefore, the staggered-flux state is a $p$-symmetry form factor density wave. Please note that a $d$-wave form factor in our notation refers to a distinct state below, which should *not* be confused with the "$d$-density wave" of Refs. [2]-[6]. With $P(\mathbf{k}, \mathbf{Q})$ non-zero only for $\mathbf{Q} = 0$ and odd in $\mathbf{k}$, we obtain states with spontaneous uniform currents [7].

Another much-studied state is the electronic nematic [8]-[10]. This has $P(\mathbf{k}, \mathbf{Q})$ non-zero only for $\mathbf{Q} = 0$, with

$$P(\mathbf{k}, \mathbf{Q}) = P_n(\cos(k_x) - \cos(k_y)) + P'_n(\cos(2k_x) - \cos(2k_y)) + \cdots \qquad (1.7)$$

Now all terms on the right-hand-side should be even in $\mathbf{k}$, and odd under the interchange $k_x \leftrightarrow k_y$. The ansatz in Eq. (1.7) also applies to "incommensurate nematics" [21]-[31] which have $P(\mathbf{k}, \mathbf{Q})$ non-zero only for $\mathbf{Q} = (\pm Q, \pm Q)$: these are density waves with $\mathbf{Q}$ along the diagonals of the square lattice Brillouin zone, and a purely $d$-wave form factor.

Finally, we turn to the density waves considered in our manuscript. These have $P(\mathbf{k}, \mathbf{Q})$ non-zero only for $\mathbf{Q} = (0, \pm Q)$ and $(\pm Q, 0)$. We assume they preserve time-reversal, and then the form-factor has the general form [22]

$$P(\mathbf{k}, \mathbf{Q}) = P_S + P_{S'}(\cos(k_x) + \cos(k_y)) + P_D(\cos(k_x) - \cos(k_y)) + \cdots \qquad (1.8)$$

For general incommensurate Q, any even function of **k** is allowed on the right-hand-side. Using arguments based upon instabilities of metals with local antiferromagnetic correlations, it was argued in Ref. [22] that such a density wave is predominantly $d$-wave, i.e., $|P_D| \gg |P_S|$ and $|P_D| \gg |P_{S'}|$, so that it is very nearly, but not exactly, an incommensurate nematic. The $d$-wave-ness here is a statement about the physics of the intra-unit-cell electronic correlations, and is not fully determined by symmetry.

We now make contact with the local observables considered in SI section II and III and as measured by STM. Via the canonical transformation from the two-band to the single-band model of the $CuO_2$ layer, we can deduce the general relationship

$$\langle c_{i\alpha}^\dagger c_{j\alpha} + c_{j\alpha}^\dagger c_{i\alpha} \rangle = \begin{cases} \frac{1}{K}\rho(\mathbf{r}_{Cu}) & \text{for } i=j \\[6pt] \frac{1}{K'}\rho(\mathbf{r}_{O_x}) & \text{for } i,j \text{ n.n along x-direction} \\[6pt] \frac{1}{K'}\rho(\mathbf{r}_{O_y}) & \text{for } i,j \text{ n.n along y-direction} \end{cases} \quad (1.9)$$

Here $\rho(\mathbf{r})$ is any density-like (i.e. invariant under time-reversal and spin rotations) observable and $K, K'$ are proportionality constants. Combining (1.1) (1.8) (1.9) we can now write

$$\begin{aligned} \rho(\mathbf{r}_{Cu}) &= 2K \mathrm{Re}\left\{ \left[ \sum_k P(\mathbf{k}, \mathbf{Q}) \right] e^{i\mathbf{Q}\cdot\mathbf{r}_{Cu}} \right\} \\ &= A_S \cos(\mathbf{Q}\cdot\mathbf{r}_{Cu} + \phi_S) \end{aligned} \quad (1.10)$$

$$\begin{aligned} \rho(\mathbf{r}_{O_x}) &= 2K' \mathrm{Re}\left\{ \left[ \sum_k \cos(k_x) P(\mathbf{k}, \mathbf{Q}) \right] e^{i\mathbf{Q}\cdot\mathbf{r}_{O_x}} \right\} \\ &= A_{S'} \cos(\mathbf{Q}\cdot\mathbf{r}_{O_x} + \phi_{S'}) + A_D \cos(\mathbf{Q}\cdot\mathbf{r}_{O_x} + \phi_D) \end{aligned} \quad (1.11)$$

$$\begin{aligned} \rho(\mathbf{r}_{O_y}) &= 2K' \mathrm{Re}\left\{ \left[ \sum_k \cos(k_y) P(\mathbf{k}, \mathbf{Q}) \right] e^{i\mathbf{Q}\cdot\mathbf{r}_{O_y}} \right\} \\ &= A_{S'} \cos(\mathbf{Q}\cdot\mathbf{r}_{O_y} + \phi_{S'}) - A_D \cos(\mathbf{Q}\cdot\mathbf{r}_{O_y} + \phi_D) \end{aligned} \quad (1.12)$$

with $A_S = 2K|P_S|$, $A_{S',D} = K'|P_{S',D}|$ and $\phi_{S,S',D} = \arg(P_{S,S',D})$. Our Fourier transforms of the STM data in Fig. 3 of the main text yield the prefactors in Eqs. (1.11) & (1.12). The observed change in sign between the prefactors demonstrates that $|P_D| \gg |P_{S'}|$ as anticipated in Refs. [22–24].

## II  Symmetry Decomposition of CuO$_2$ IUC States

Here we present mathematical details behind the angular momentum form factor organization of density waves on the CuO$_2$ plane. Among the many ways of organizing density waves in the CuO$_2$ plane, one is to think of them as a wave on the copper atoms, a wave on the x-axis bond oxygen atoms and a wave on the y-axis bond oxygen atoms as presented in Eqs. (1.10)-(1.12).

The results presented in the main manuscript, however, present a compelling case that another organization captures the density wave observed *in a remarkably simple way*. This way organizes them by angular momentum form factors that we call $s$, $s'$ ("extended $s$") and $d$.

We can think of the angular momentum form factor organization as a modulation of $\mathbf{Q} = 0$ "waves" whose point group symmetry is well defined, as shown in Figs. S1A,B,C. The $\mathbf{Q} = 0$ $s$-wave has a density

$$\rho(\mathbf{r}_{Cu}) = A_S, \ \rho(\mathbf{r}_{O_x}) = 0, \ \rho(\mathbf{r}_{O_y}) = 0, \tag{2.1}$$

the $\mathbf{Q} = 0$ $s'$-wave has density

$$\rho(\mathbf{r}_{Cu}) = 0, \ \rho(\mathbf{r}_{O_x}) = A_{S'}, \ \rho(\mathbf{r}_{O_y}) = A_{S'}, \tag{2.2}$$

and the $\mathbf{Q} = 0$ $d$-wave has density

$$\rho(\mathbf{r}_{Cu}) = 0, \ \rho(\mathbf{r}_{O_x}) = A_D, \ \rho(\mathbf{r}_{O_y}) = -A_D. \tag{2.3}$$

The Fourier transforms of these IUC states are shown in Figs. S1D,E,F. Modulating these waves, we then obtain

$$\rho_S(\mathbf{r}) = \begin{cases} A_S \cos(\mathbf{Q} \cdot \mathbf{r} + \phi_S), & \mathbf{r} = \mathbf{r}_{Cu}, \\ 0, & \mathbf{r} = \mathbf{r}_{O_x}, \\ 0, & \mathbf{r} = \mathbf{r}_{O_y}, \end{cases}$$

$$\rho_{S'}(\mathbf{r}) = \begin{cases} 0, & \mathbf{r} = \mathbf{r}_{Cu}, \\ A_{S'} \cos(\mathbf{Q} \cdot \mathbf{r} + \phi_{S'}), & \mathbf{r} = \mathbf{r}_{O_x}, \\ A_{S'} \cos(\mathbf{Q} \cdot \mathbf{r} + \phi_{S'}), & \mathbf{r} = \mathbf{r}_{O_y}, \end{cases}$$

$$\rho_D(\mathbf{r}) = \begin{cases} 0, & \mathbf{r} = \mathbf{r}_{Cu}, \\ A_D \cos(\mathbf{Q} \cdot \mathbf{r} + \phi_D), & \mathbf{r} = \mathbf{r}_{O_x}, \\ -A_D \cos(\mathbf{Q} \cdot \mathbf{r} + \phi_D), & \mathbf{r} = \mathbf{r}_{O_y}, \end{cases} \qquad (2.4)$$

where $\phi_{S,S',D}$ are the phases of each of the DW form factor components. A graphical picture corresponding to these waves is presented in Figs. S2A,B,C. The Fourier transforms of the three different form factor density waves are presented in Figs. S2D,E,F and will be considered further in SI section III. If, for simplicity, we choose $\phi_S = \phi_{S'} = \phi_D = \phi(\mathbf{r})$ and allow for spatial disorder of the phase we arrive at the description used in eq. (1) of the main text.

Consider now the organization by atomic site. We see that the $s$-wave form factor is just a wave purely on the copper atoms with no weight on the oxygen atoms while the $s'$-wave and $d$-wave form factors involve purely the oxygen sites. There is also a curious but practically very important relationship between the $s'$-wave and $d$-wave form factors: in a sense they are like mirror images of each other. For a purely $s'$-form factor DW, taking the sum $\tilde{\rho}_{O_x}(\mathbf{q}) + \tilde{\rho}_{O_y}(\mathbf{q})$ must recover the Fourier transform of the full $s'$-form factor DW. However, taking the difference $\tilde{\rho}_{O_x}(\mathbf{q}) - \tilde{\rho}_{O_y}(\mathbf{q})$, we obtain the Fourier transform of the $d$-form factor DW (up to a phase difference $\phi_S - \phi_D$). Similarly, for a density wave with a pure $d$-symmetry form factor, $\tilde{\rho}_{O_x}(\mathbf{q}) - \tilde{\rho}_{O_y}(\mathbf{q})$ will look like the Fourier transform of a DW with a pure $s'$-symmetry form factor.

Finally, given the above understanding of how the overall electronic structure image (e.g. $R(\mathbf{r})$) is built up from its components, there is another possible approach to determining the form factor of any density wave. Phase-resolved Fourier analysis of such an electronic structure image that has not been decomposed into its constituent parts $Cu(\mathbf{r})$, $O_x(\mathbf{r})$, $O_y(\mathbf{r})$ but remains intact, should still reveal the relative magnitude of the three form factors. However, one can show that this is only possible if the three independent DW peaks at $\mathbf{Q}$, $\mathbf{Q}' = (1,0) + \mathbf{Q}$ and $\mathbf{Q}'' = (0,1) + \mathbf{Q}$ are well resolved.

## III  Predicted Fourier Transform STM Signatures of a dFF DW

As discussed in SI sections I and II, the projection of a density wave (DW) into $s$, $s'$ and $d$ form factor components is conceptually appealing. However, for the purposes of this section we will keep in mind the exigencies of the experimental technique and work in terms of the segregated oxygen sub-lattice images $O_{x,y}(\mathbf{r})$. In terms of the segregated sub-lattices, a $d$-form factor DW is one for which the DW on the $O_x$ sites is in anti-phase with that on the $O_y$ sites. For $\mathbf{Q} \neq 0$ ordering the form factor does not uniquely determine the point group symmetry of the DW and hence in general $s$, $s'$ and $d$ form factors are free to mix. This section predicts the consequences of a primarily $d$-form factor density wave for $\tilde{O}_{x,y}(\mathbf{q})$ and shows its consistency with the data presented in the main text.

To deduce the logical consequences of a $d$-form factor DW for the Fourier transforms of the segregated oxygen site images one can start by constructing the dual real and momentum-space representation of the sub-lattices:

$$L_{Cu}(\mathbf{r}) = \sum_{i,j} \delta(\mathbf{r} - \mathbf{R}_{i,j}) \Leftrightarrow \tilde{L}_{Cu}(\mathbf{q}) = \sum_{h,k} \delta(\mathbf{q} - \mathbf{G}^{h,k}) \tag{3.1}$$

$$L_{O_x}(\mathbf{r}) = L_{Cu}(\mathbf{r} - \frac{a_0 \hat{x}}{2}) \Leftrightarrow \tilde{L}_{O_x}(\mathbf{q}) = e^{i\mathbf{q}\cdot(a_0 \hat{x}/2)} \tilde{L}_{Cu}(\mathbf{q}) \tag{3.2}$$

$$L_{O_y}(\mathbf{r}) = L_{Cu}(\mathbf{r} - \frac{a_0 \hat{y}}{2}) \Leftrightarrow \tilde{L}_{O_y}(\mathbf{q}) = e^{i\mathbf{q}\cdot(a_0 \hat{y}/2)} \tilde{L}_{Cu}(\mathbf{q}) \tag{3.3}$$

The $\{\mathbf{R}_{i,j}\}$ are the set of direct lattice vectors of the square lattice with lattice constant $a_0$ and the $\{\mathbf{G}^{h,k}\}$ are the reciprocal lattice vectors. The displacement of the oxygen sub-lattices from the copper sub-lattice has the effect of shifting the phase of their Bragg peaks along the direction of displacement by $\pi$ in reciprocal space. This is depicted in Fig. S3A.

Using the convolution theorem, a $d$-form factor modulation of the oxygen site density takes on the dual description:

$$O_x(\mathbf{r}) = L_{O_x}(\mathbf{r}) \cdot A_{O_x}(\mathbf{r}) \Leftrightarrow \tilde{O}_x(\mathbf{q}) = \tilde{L}_{O_x}(\mathbf{q}) * \tilde{A}_{O_x}(\mathbf{q}) \tag{3.4}$$

$$O_y(\mathbf{r}) = L_{O_y}(\mathbf{r}) \cdot A_{O_y}(\mathbf{r}) \Leftrightarrow \tilde{O}_y(\mathbf{q}) = \tilde{L}_{O_y}(\mathbf{q}) * \tilde{A}_{O_y}(\mathbf{q}) \tag{3.5}$$

$$A_{O_x}(\mathbf{r}) = -A_{O_y}(\mathbf{r}) = A(\mathbf{r}) \Leftrightarrow \tilde{A}_{O_x}(\mathbf{q}) = -\tilde{A}_{O_y}(\mathbf{q}) = \tilde{A}(\mathbf{q}) \tag{3.6}$$

The functions $O_{x,y}(\mathbf{r})$ are the segregated oxygen sub-lattice images. The $A_{O_{x,y}}(\mathbf{r})$ are continuous functions that when multiplied by the sub-lattice functions yield density waves in anti-phase on the separate oxygen sub-lattices (Fig. S3B). Fig. S3C shows their Fourier transforms $\tilde{A}_{O_{x,y}}(\mathbf{q})$. Note that $A(\mathbf{r})$ may contain arbitrary amplitude and overall phase disorder but remain $d$-wave so long as the relative phase relation in Eq. (3.6) is maintained.

As shown in Fig. S4A, the convolutions in Eqs. (3.4) & (3.5) create an image of $\tilde{A}_{O_{x,y}}(\mathbf{q})$ at each reciprocal lattice vector that sum to form the total convolution. Labelling the convolution image due to the reciprocal lattice vector $(h,k)$ in the x sub-lattice $\tilde{O}_x^{h,k}(\mathbf{q})$:

$$\tilde{O}_x(\mathbf{q}) = \sum_{h,k} \tilde{O}_x^{h,k}(\mathbf{q}) = \sum_{h,k} e^{i\mathbf{G}^{h,k}\cdot(a_0\hat{x}/2)} \tilde{A}_{O_x}(\mathbf{q} - \mathbf{G}^{h,k}) \qquad (3.7)$$

In creating the $(h,k)$ convolution image, the phase of the sub-lattice Bragg peak at $\mathbf{G}^{h,k}$ and that of the form factor $\tilde{A}_{O_x}(\mathbf{q})$ must be added:

$$\arg\left\{\tilde{O}_x^{h,k}(\mathbf{q})\right\} = \arg\left\{\tilde{A}_{O_x}(\mathbf{q} - \mathbf{G}^{h,k})\right\} + \arg\left\{e^{i\mathbf{G}^{h,k}\cdot(a_0\hat{x}/2)}\right\} \qquad (3.8)$$

Thus it follows immediately that

$$\tilde{O}_x^{0,0} = A(\mathbf{q}),\ \tilde{O}_y^{0,0} = -A(\mathbf{q}) \qquad (3.9)$$

$$\tilde{O}_x^{1,0} = -A(\mathbf{q}),\ \tilde{O}_y^{1,0} = -A(\mathbf{q}) \qquad (3.10)$$

$$\tilde{O}_x^{0,1} = A(\mathbf{q}),\ \tilde{O}_y^{0,1} = A(\mathbf{q}) \qquad (3.11)$$

and hence

$$\tilde{O}_x^{0,0} + \tilde{O}_y^{0,0} = 0,\ \tilde{O}_x^{0,0} - \tilde{O}_y^{0,0} = 2A(\mathbf{q}) \qquad (3.12)$$

$$\tilde{O}_x^{1,0} + \tilde{O}_y^{1,0} = -2A(\mathbf{q}),\ \tilde{O}_x^{1,0} - \tilde{O}_y^{1,0} = 0 \qquad (3.13)$$

$$\tilde{O}_x^{0,1} + \tilde{O}_y^{0,1} = 2A(\mathbf{q}),\ \tilde{O}_x^{0,1} - \tilde{O}_y^{0,1} = 0 \qquad (3.14)$$

A direct consequence of a $d$-form factor is that in $\tilde{O}_x(\mathbf{q}) + \tilde{O}_y(\mathbf{q})$ the amplitude of the convolution image at (0,0) is cancelled exactly whereas those at the $(\pm 1,0)$ and $(0,\pm 1)$ points are enhanced as illustrated in Figs. S4B & C. The converse is true for $\tilde{O}_x(\mathbf{q}) - \tilde{O}_y(\mathbf{q})$. This holds for any $d$-wave modulation in the presence of arbitrary amplitude and overall phase disorder.

Figs. S2D-F show Fourier transforms of different form factor density waves in the $CuO_2$ plane. A $d$-form factor density wave has modulations only on the oxygen sites and hence its contribution to the full Fourier transform is contained entirely within $\tilde{O}_x(\mathbf{q}) + \tilde{O}_y(\mathbf{q})$. From Eqs. (3.12)-(3.14) we must conclude that for density waves with principal wave-vectors that lie within the 1st Brillouin zone, $\tilde{\rho}_D(\mathbf{q})$ (Fig. S2F) will exhibit an absence of peaks at these wave-vectors in the 1st Brillouin zone. For $\tilde{\rho}_S(\mathbf{q})$ (Fig. S2D) and $\tilde{\rho}_{S'}(\mathbf{q})$ (Fig. S2E) we may conclude that they will be present using similar arguments.

Empirically (main text Figs. 3 and 4), our data contain modulations at two wavevectors $\mathbf{Q}_x = (Q,0)$ and $\mathbf{Q}_y = (0,Q)$ with Q$\approx$1/4 but with a great deal of fluctuation in the spatial-phase of the DW (see Ref. 43 of main text). However, it would be improper to conclude from this that we observe a bi-directional $d$-form factor DW, often termed the "checkerboard" modulation. The strong disorder of the density modulations in BSCCO and NaCCOC is apparent in the real-space images presented in Fig. 2 of the main text and Section V of this document. Random charge disorder can have the effect of taking a clean system with an instability toward uni-directional ("stripe") ordering and produce domains of uni-directional order that align with the local anisotropy. Conversely, a clean system with an instability towards bi-directional ("checkerboard") ordering may have local anisotropy imbued upon it by disorder.

Whilst the wave-vector(s) of the underlying instability of the copper oxide plane to DW ordering are of theoretical interest, any $d$-form factor DW containing two wave-vectors $\mathbf{Q}_x$ and $\mathbf{Q}_y$ can be described by:

$$A(\mathbf{r}) = \text{Re}[e^{i\mathbf{Q}_x \cdot \mathbf{r}} H_x(\mathbf{r})] + \text{Re}[e^{i\mathbf{Q}_y \cdot \mathbf{r}} H_y(\mathbf{r})] \tag{3.15}$$

$$A(\mathbf{q}) = \frac{1}{2}\left[\tilde{H}_x(\mathbf{q}-\mathbf{Q}_x) + \tilde{H}_x^*(\mathbf{q}+\mathbf{Q}_x) + \tilde{H}_y(\mathbf{q}-\mathbf{Q}_y) + \tilde{H}_y^*(\mathbf{q}+\mathbf{Q}_y)\right]. \tag{3.16}$$

The complex valued functions $H_{x,y}(\mathbf{r})$ locally modulate the amplitude and phase of the density wave and hence encode its disorder. The problem now reduces to performing the convolutions contained in Eqs. (3.4), (3.5) & (3.6).

For the specific example of $\mathbf{Q}_x \approx (1/4, 0)$ and $\mathbf{Q}_y \approx (0, 1/4)$ considered in our study the primarily $d$-wave form factor requires that the peaks at $\pm\mathbf{Q}_x$ and $\pm\mathbf{Q}_y$ present in both $\tilde{O}_x(\mathbf{q})$ and $\tilde{O}_y(\mathbf{q})$ must cancel exactly in $\tilde{O}_x(\mathbf{q}) + \tilde{O}_y(\mathbf{q})$ and be enhanced in $\tilde{O}_x(\mathbf{q}) - \tilde{O}_y(\mathbf{q})$. Conversely the peaks at $\mathbf{Q}' = (1,0) \pm \mathbf{Q}_{x,y}$ and $\mathbf{Q}'' = (0,1) \pm \mathbf{Q}_{x,y}$ will be enhanced in $\tilde{O}_x(\mathbf{q}) + \tilde{O}_y(\mathbf{q})$ but will cancel exactly in $\tilde{O}_x(\mathbf{q}) - \tilde{O}_y(\mathbf{q})$. These are necessary consequences of a DW with a primarily $d$-wave form factor. This is discussed in the main text and in accord with the observations in Figs. 2-4 of the main text.

## IV  Sublattice Phase Definition: Lawler-Fujita Algorithm

Consider an atomically resolved STM topograph, $T(\mathbf{r})$, with tetragonal symmetry where two orthogonal wavevectors generate the atomic corrugations. These are centered at the first reciprocal unit cell Bragg wavevectors $\mathbf{Q}_a = (Q_{ax}, Q_{ay})$ and $\mathbf{Q}_b = (Q_{bx}, Q_{by})$ with $\mathbf{a}$ and $\mathbf{b}$ being the unit cell vectors. Schematically, the ideal topographic image can be written as

$$T(\mathbf{r}) = T_0[\cos(\mathbf{Q}_a \cdot \mathbf{r}) + \cos(\mathbf{Q}_b \cdot \mathbf{r})]. \tag{4.1}$$

In SI-STM, the $T(\mathbf{r})$ and its simultaneously measured spectroscopic current map, $I(\mathbf{r}, V)$, and differential conductance map, $g(\mathbf{r}, V)$, are specified by measurements on a square array of pixels with coordinates labeled $\mathbf{r} = (x, y)$. The power-spectral-density (PSD) Fourier transform of $T(\mathbf{r})$, $|\tilde{T}(\mathbf{q})|^2$ – where $\tilde{T}(\mathbf{q}) = \mathrm{Re}\tilde{T}(\mathbf{q}) + i\mathrm{Im}\tilde{T}(\mathbf{q})$, then exhibits two distinct peaks at $\mathbf{q} = \mathbf{Q}_a$ and $\mathbf{Q}_b$.

In an actual experiment, $T(\mathbf{r})$ suffers picometer scale disortions from the ideal representation in (4.1) according to a slowly varying 'displacement field', $\mathbf{u}(\mathbf{r})$. The same distortion is also found in the spectroscopic data. Thus, a topographic image, including distortions, is schematically written as

$$T(\mathbf{r}) = T_0[\cos(\mathbf{Q}_a \cdot (\mathbf{r} + \mathbf{u}(\mathbf{r}))) + \cos(\mathbf{Q}_b \cdot (\mathbf{r} + \mathbf{u}(\mathbf{r})))]. \tag{4.2}$$

Then, to remove the effects of $\mathbf{u}(\mathbf{r})$ requires an affine transformation at each point in space.

To begin, define the local phase of the atomic cosine components, at a given point $\mathbf{r}$, as

$$\begin{aligned} \varphi_a(\mathbf{r}) &= \mathbf{Q}_a \cdot \mathbf{r} + \theta_a(\mathbf{r}) \\ \varphi_b(\mathbf{r}) &= \mathbf{Q}_b \cdot \mathbf{r} + \theta_b(\mathbf{r}) \end{aligned} \tag{4.3}$$

which recasts equation (4.2) as

$$T(\mathbf{r}) = T_0[\cos\left(\varphi_a(\mathbf{r})\right) + \cos\left(\varphi_b(\mathbf{r})\right)] \tag{4.4}$$

where $\theta_i(\mathbf{r}) = \mathbf{Q}_i \cdot \mathbf{u}(\mathbf{r})$ is additional phase generated by the displacement field. If there were no distortions and the $T(\mathbf{r})$ image were perfectly periodic then $\theta_i(\mathbf{r})$ would be constant. From this perspective, the 2-dimensional lattice in (4.4) is a function of phase alone. For example, the apex of every atom in the topographic image has the same phase, $0 \pmod{2\pi}$ regardless of where it is in the image. When viewed in the $\mathbf{r}$ coordinates, the distance between such points of equal phase in the 'perfect' lattice and distorted lattice is not the same. The problem of correcting $T(\mathbf{r})$ then reduces to finding a transformation to map the distorted lattice onto the 'perfect' one, using the phase information $\varphi_i(\mathbf{r})$. This is equivalent to finding a set of local transformations which makes $\theta_a(\mathbf{r})$ and $\theta_b(\mathbf{r})$ constant over all space; call them $\bar{\theta}_a$ and $\bar{\theta}_b$ respectively.

Let $\mathbf{r}$ be a point on the unprocessed $T(\mathbf{r})$ and let $\tilde{\mathbf{r}}$ be the point of equal phase on the perfect lattice periodic image, which needs to be determined. This produces a set of equivalency relations

$$\mathbf{Q}_a \cdot \mathbf{r} + \theta_a(\mathbf{r}) = \mathbf{Q}_a \cdot \tilde{\mathbf{r}} + \bar{\theta}_a$$
$$\mathbf{Q}_b \cdot \mathbf{r} + \theta_b(\mathbf{r}) = \mathbf{Q}_b \cdot \tilde{\mathbf{r}} + \bar{\theta}_b$$
(4.5)

Solving for $\tilde{\mathbf{r}} = (\tilde{x}, \tilde{y})$ and then assigning the values of the topographic image at $\mathbf{r} = (x, y)$, $T(\mathbf{r})$, to $\tilde{\mathbf{r}}$ produces the 'perfect' lattice. To solve for $\tilde{\mathbf{r}}$ rewrite (4.5) in matrix form

$$\mathbf{Q} \begin{pmatrix} \tilde{x} \\ \tilde{y} \end{pmatrix} = \mathbf{Q} \begin{pmatrix} x \\ y \end{pmatrix} - \begin{pmatrix} \bar{\theta}_a - \theta_a(\mathbf{r}) \\ \bar{\theta}_b - \theta_b(\mathbf{r}) \end{pmatrix}$$
(4.6)

where

$$\mathbf{Q} = \begin{pmatrix} Q_{ax} & Q_{ay} \\ Q_{bx} & Q_{by} \end{pmatrix}.$$
(4.7)

Because $\mathbf{Q}_a$ and $\mathbf{Q}_b$ are orthogonal, $\mathbf{Q}$ is invertible allowing one to solve for the displacement field $\mathbf{u}(\mathbf{r})$ which maps $\mathbf{r}$ to $\tilde{\mathbf{r}}$:

$$\mathbf{u}(\mathbf{r}) = \mathbf{Q}^{-1} \begin{pmatrix} \bar{\theta}_a - \theta_a(\mathbf{r}) \\ \bar{\theta}_b - \theta_b(\mathbf{r}) \end{pmatrix}.$$
(4.8)

In practice, we use the convention $\bar{\theta}_i = 0$ which generates a 'perfect' lattice with an atomic peak centered at the origin. This is equivalent to setting to zero the imaginary component of the Bragg peaks in the Fourier transform.

Of course, to employ the transformation in (4.6) one must first extract $\theta_i(\mathbf{r})$ from the topographic data. This is accomplished by using a computational lock-in technique in which the topograph, $T(\mathbf{r})$, is multiplied by reference sine and cosine functions with periodicity set by $\mathbf{Q}_a$ and $\mathbf{Q}_b$. The resulting four images are filtered to retain only the $q$-space regions within a radius $\delta q = 1/\lambda$ of the four Bragg peaks; the magnitude of $\lambda$ is chosen to capture only the relevant image distortions. This procedure results in retaining the local phase information $\theta_a(\mathbf{r})$, $\theta_b(\mathbf{r})$ that quantifies the local displacements from perfect periodicity:

$$Y_i(\mathbf{r}) = \sin \theta_i(\mathbf{r}), \ X_i(\mathbf{r}) = \cos \theta_i(\mathbf{r})$$
(4.9)

Dividing the appropriate pair of images allows one to extract $\theta_i(\mathbf{r})$:

$$\theta_i(\mathbf{r}) = \tan^{-1} \frac{Y_i(\mathbf{r})}{X_i(\mathbf{r})}. \tag{4.10}$$

## V   Data Analysis

In Fig. S5 we show the power spectral density Fourier transform analysis $|\tilde{C}u(\mathbf{q})|^2$ and $|[\tilde{O}_x(\mathbf{q})+\tilde{O}_y(\mathbf{q})]/2|^2$ yielding the $s$-form factor magnitude $S$ and the $s'$-form factor magnitude $S'$, respectively. These are the equivalent result for $|[\tilde{O}_x(\mathbf{q}) - \tilde{O}_y(\mathbf{q})]/2|^2$ which is shown in the Fig. 4A of the main text. The measured values $S$, $S'$ are plotted along the dashed lines through $\mathbf{Q}$ together with the value $D$ in Fig. 4B of the main text.

In Fig. 4D of main text, we show the 2D histogram of the amplitude difference and the phase difference between $O_x(\mathbf{r})$ and $O_y(\mathbf{r})$. In order to construct this, first, the magnitude and the phase only associated with $\mathbf{Q}_x \sim (1/4, 0)$ and $\mathbf{Q}_y \sim (0, 1/4)$ are calculated by using the fourier filtration in $O_x(\mathbf{r})$ and $O_y(\mathbf{r})$.

$$\tilde{O}_\alpha(\mathbf{r}, \mathbf{q}) = \int d\mathbf{R} O_\alpha(\mathbf{R}) e^{i\mathbf{q}\cdot\mathbf{R}} e^{-\frac{|\mathbf{r}-\mathbf{R}|^2}{2\Lambda^2}} \frac{1}{2\pi\Lambda^2}, \tag{5.1}$$

where $\alpha, \beta = x, y$ and $\Lambda$ the averaging length to be $\sim 30$Å.

For $\mathbf{q} = \mathbf{Q}_\beta$, amplitudes and phases are given by

$$|\tilde{O}_\alpha(\mathbf{r}, \mathbf{Q}_\beta)| = \sqrt{\text{Re}\tilde{O}_\alpha(\mathbf{r}, \mathbf{Q}_\beta)^2 + \text{Im}\tilde{O}_\alpha(\mathbf{r}, \mathbf{Q}_\beta)^2}, \tag{5.2}$$

$$\arg[\tilde{O}_\alpha(\mathbf{r}, \mathbf{Q}_\beta)] = \tan^{-1} \frac{\text{Im}\tilde{O}_\alpha(\mathbf{r}, \mathbf{Q}_\beta)}{\text{Re}\tilde{O}_\alpha(\mathbf{r}, \mathbf{Q}_\beta)}. \tag{5.3}$$

Next, the normalized amplitude difference and the phase difference are then defined by

$$\frac{|\tilde{O}_x(\mathbf{r}, \mathbf{Q}_\beta)| - |\tilde{O}_y(\mathbf{r}, \mathbf{Q}_\beta)|}{|\tilde{O}_x(\mathbf{r}, \mathbf{Q}_\beta)| + |\tilde{O}_y(\mathbf{r}, \mathbf{Q}_\beta)|}, \tag{5.4}$$

$$\left|\arg[\tilde{O}_x(\mathbf{r}, \mathbf{Q}_\beta)] - \arg[\tilde{O}_y(\mathbf{r}, \mathbf{Q}_\beta)]\right|. \tag{5.5}$$

Finally, using (5.4) and (5.5) we obtain a two dimensional histogram for both $\mathbf{Q}_x$ and $\mathbf{Q}_y$, independently, and then take sum of them to construct single distribution containing the information for both $\mathbf{Q}_x$ and $\mathbf{Q}_y$.

In Fig. S6 we show the measured $R(\mathbf{r})$ (subset of main Fig. 2A is presented since original FOV is so large that the DW is no longer visible clearly) and its segregation into three site-specific images $Cu(\mathbf{r})$, $O_x(\mathbf{r})$ and $O_y(\mathbf{r})$ as described in the main text. With the origin set at a Cu site, Fig. S7 then shows the three complex valued Fourier transform images derived from Fig. 2A: $\tilde{C}u(\mathbf{q}) \equiv \text{Re}\tilde{C}u(\mathbf{q}) + i\text{Im}\tilde{C}u(\mathbf{q})$, $\tilde{O}_x(\mathbf{q}) \equiv \text{Re}\tilde{O}_x(\mathbf{q}) + i\text{Im}\tilde{O}_x(\mathbf{q})$, $\tilde{O}_y(\mathbf{q}) \equiv \text{Re}\tilde{O}_y(\mathbf{q}) + i\text{Im}\tilde{O}_y(\mathbf{q})$. This type of sublattice-phase-resolved Fourier analysis which we introduce in this paper provides the capability to measure the relative phase of different sites with each CuO$_2$ unit cell. The inset to Fig. S6A shows the difference between the real component of Bragg intensity for (1,0) and (0,1) peaks in the Fourier transforms of the electronic structure images *before* sublattice segregation. It is this difference that was originally used to determine the *d*-form factor of the intra-unit-cell nematic state; see Ref. 12, 44 of main text. Figures S8 and S9 present the corresponding data and analysis for NaCCOC.

Figure S10 compares the analysis of $Z(\mathbf{r},|E|) = g(\mathbf{r},E)/g(\mathbf{r},-E)$ (E=150meV) between BSCCO (S10A-D) and NaCCOC (S10E-H). Both $Z(\mathbf{r},|E|)$ are segregated into three site-specific images $Cu(\mathbf{r})$, $O_x(\mathbf{r})$ and $O_y(\mathbf{r})$ first. The analysis is then presented in terms of their complex Fourier transforms $\mathrm{Re}\tilde{O}_x(\mathbf{q})$, $\mathrm{Re}\tilde{O}_y(\mathbf{q})$ as described in the main text. One can see directly that the phenomena are extremely similar for both compounds, in terms of $\mathrm{Re}\tilde{O}_x(\mathbf{q})$, $\mathrm{Re}\tilde{O}_y(\mathbf{q})$ and $\mathrm{Re}\tilde{O}_x(\mathbf{q}) \pm \mathrm{Re}\tilde{O}_y(\mathbf{q})$. Moreover they are in excellent agreement with expectations for a dFF-DW in Figs S4B,C. Thus, in the main text, we present analysis of $Z(\mathbf{r}, E = 150meV)$ on an equivalent basis to $R(\mathbf{r}, E = 150meV)$ when deriving $\tilde{O}_x(\mathbf{q}) \equiv \mathrm{Re}\tilde{O}_x(\mathbf{q}) + i\mathrm{Im}\tilde{O}_x(\mathbf{q})$, $\tilde{O}_y(\mathbf{q}) \equiv \mathrm{Re}\tilde{O}_y(\mathbf{q}) + i\mathrm{Im}\tilde{O}_y(\mathbf{q})$ for Fig. 3E-H of the main text.

## Supporting Figure Captions

**Figure S1 Intra-unit-cell Electronic Structure Symmetry in the CuO$_2$ Plane**

A. Schematic of uniform density on the Cu atoms ($s$-symmetry). The inactive O sites are now indicated by black dots.

B. Schematic of uniform density on the O atoms (also an $s$-symmetry referred to here as extended-$s$ or $s'$-symmetry). The inactive Cu sites are indicated by black dots.

C. Schematic pattern with opposite-sign density on $O_x$ and $O_y$ ($d$-symmetry). The inactive Cu sites are indicated by black dots.

D. Real component of Fourier transform of the $s$-symmetry IUC patterns derived only from Cu sublattice in (A) and with no DW modulation. The Bragg peaks have the same sign indicating the IUC states have $s$-symmetry.

E. Real component of Fourier transform of the $s'$-symmetry IUC patterns derived only from $O_x$ and $O_y$ sublattices in (B) and with no DW. The Bragg peaks are no longer within the CuO$_2$ reciprocal unit cell (RUC).

F. Real component of Fourier transform of the $d$-symmetry IUC patterns derived only from $O_x$ and $O_y$ sublattices as shown in (C) and with no DW modulation. The Bragg peaks now have the opposite sign indicating the IUC states have $d$-symmetry.

**Figure S2 Types of CuO$_2$ Intra-unit-cell Density Waves**

A. Spatial modulation with wavevector $\mathbf{Q} = (Q,0)$ of the $s$-symmetry IUC patterns in (S1A) is described by $\rho_S(\mathbf{r}) = S(\mathbf{r})\cos(\mathbf{Q}\cdot\mathbf{r})$; only Cu sites are active. The inactive O sites indicated by black dots.

B. Spatial modulation with wavevector $\mathbf{Q}$ of the patterns in (S1B) described by $\rho_{S'}(\mathbf{r}) = S'(\mathbf{r})\cos(\mathbf{Q}\cdot\mathbf{r})$; only $O_x$ and $O_y$ sites are active but they are always equivalent within each unit cell. The inactive Cu sites are indicated by black dots.

C. Spatial modulation with wavevector $\mathbf{Q}$ of the patterns in (S1C) described by $\rho_D(\mathbf{r}) = D(\mathbf{r})\cos(\mathbf{Q}\cdot\mathbf{r})$; only $O_x$ and $O_y$ sites are relevant but now they are always inequivalent and indeed $\pi$ out of phase.

D. Re$\tilde{\rho}_S(\mathbf{q})$, the real-component of Fourier transform of the pattern in (S2A). For this $s$-form factor DW, the DW satellites of inequivalent Bragg peaks $\mathbf{Q}'$ and $\mathbf{Q}''$ exhibit same sign.

E. Re$\tilde{\rho}_{S'}(\mathbf{q})$, the real-component of Fourier transform of the pattern in (S2B). For this $s'$-form factor DW, the peaks at $\mathbf{Q}$ are clear and the actual Bragg peaks of (S2B) are outside the RUC of $CuO_2$.

F. Re$\tilde{\rho}_D(\mathbf{q})$, the real-component of Fourier transform of the pattern in (S2C). For this $d$-form factor DW, the DW Bragg-satellite peaks at $\mathbf{Q}'$ and $\mathbf{Q}''$ exhibit opposite sign. More profoundly, because they are out of phase by $\pi$ the contributions of $O_x$ and $O_y$ sites in each unit cell cancel, resulting in the disappearance of the DW modulation peaks $\mathbf{Q}$ within the BZ (dashed box).

**Figure S3 Sub-lattice Decomposition of $d$ Form Factor DW**

A. Fourier transforms of the x-bond and y-bond oxygen sublattices without a DW modulation. Grey orbitals signify those in the sub-lattice under consideration and black those in the other sub-lattice.

B. Schematic of continuous functions $A_{O_{x,y}}(\mathbf{r})$ which when multiplied by the sublattice functions $L_{O_{x,y}}(\mathbf{r})$ yield density waves in anti-phase on the two sublattices with a modulation along the x direction.

C. Fourier transforms of the functions $A_{O_{x,y}}(\mathbf{r})$ exhibiting a relative phase of $\pi$ as required for a $d$-form factor density wave.

**Figure S4 Fourier Analysis of DW using the Convolution Theorem**

A. Schematic of the segrated sublattice images $O_{x,y}(\mathbf{r})$ and their Fourier transforms $\tilde{O}_{x,y}(\mathbf{q})$ which can be obtained from Fig. S3 by application of the convolution theorem.

B. Sum and difference of Re$\tilde{O}_x(\mathbf{q})$ and Re$\tilde{O}_y(\mathbf{q})$ for a $d$-form factor density wave with modulation along the x direction at $\mathbf{Q} = (Q, 0)$. Note that the origin of co-ordinates in real space has been chosen such that the Fourier transforms are purely real.

C. Sum and difference of Re$\tilde{O}_x(\mathbf{q})$ and Re$\tilde{O}_y(\mathbf{q})$ for a $d$-form factor density wave with modulations along the x and y directions at $\mathbf{Q} = (Q, 0), (0, Q)$. The key signature of the $d$-form factor is the absence of the peaks at $(Q,0), (0,Q)$ in $\text{Re}\tilde{O}_x(\mathbf{q}) + \text{Re}\tilde{O}_y(\mathbf{q})$ and their presence in $\text{Re}\tilde{O}_x(\mathbf{q}) - \text{Re}\tilde{O}_y(\mathbf{q})$; the converse being true for the DW peaks surrounding $(\pm 1, 0)$ and $(0, \pm 1)$.

**Figure S5 Measurement of $s$-symmetry and $s'$-symmetry form factors**

A. PSD Fourier transform of $R(\mathbf{r})$ measured only at $Cu$ sites yielding $|\tilde{Cu}(\mathbf{q})|^2$. This provides the measure of relative strength of the $s$-form factor in the DW.

B. PSD Fourier transform of $R(\mathbf{r})$ measured only at $O_x/O_y$ sites yielding $|[\tilde{O}_x(\mathbf{q}) + \tilde{O}_y(\mathbf{q})]/2|^2$. This provides the measure of relative strength of the $s'$-form factor in the DW.

**Figure S6 Sublattice Segregation for BSCCO**

A. Measured $R(\mathbf{r})$ for BSCCO sample with $p \sim 8 \pm 1\%$. This data is a subset of Fig. 2A reproduced here for clarity. The inset demonstrates an inequivalence between the real component of Bragg intensity for (1,0) and (0,1) peaks in the Fourier transforms of the electronic structure image *before* sublattice segregation signalling a $\mathbf{Q} = 0$ nematic state.

B. Copper site segregated image, $Cu(\mathbf{r})$, in which the spatial average is subracted, with copper sites selected from A.

C. x-bond oxygen-site segregated image, $O_x(\mathbf{r})$, in which the spatial average is subracted, with x-oxygen sites selected from A.

D. y-bond oxygen-site segregated image, $O_y(\mathbf{r})$, in which the spatial average is subracted, with y-oxygen sites selected from A.

**Figure S7 Sublattice Phase Resolved Fourier Analysis for BSCCO**

A. Measured $\text{Re}\tilde{Cu}(\mathbf{q})$ for BSCCO sample in Fig. 2A. No DW peaks are discernable at $\mathbf{Q} = (Q, 0), (0, Q)$ or at Bragg satellites surrounding $(\pm 1, 0)$ and $(0, \pm 1)$. This indicates a very small $s$-wave component for the density wave form factor.

B. Measured $\text{Im}\tilde{Cu}(\mathbf{q})$ which also indicates a very small $s$ wave component.

C. Measured $\text{Re}\tilde{O}_x(\mathbf{q})$ showing DW peaks at $\mathbf{Q} = (Q, 0), (0, Q)$ and corresponding Bragg satellites.

D. Measured $\text{Im}\tilde{O}_x(\mathbf{q})$ which exhibits the same structure as C. The strong overall phase disorder is apparent in the colour variation within the DW peaks.

E. Measured $\text{Re}\tilde{O}_y(\mathbf{q})$ which also shows DW peaks at $\mathbf{Q} = (Q, 0), (0, Q)$ along with Bragg satellites.

F. Measured $\text{Im}\tilde{O}_y(\mathbf{q})$ which exhibits the same structure as E.

**Figure S8 Sub-Lattice Segregation for NaCCOC**

A. Measured $Z(\mathbf{r}, E = 150mV)$ for NaCCOC sample with $p \sim 12 \pm 1\%$. The inset demonstrates an inequivalence between the real component of Bragg intensity for (1,0) and (0,1) peaks in the Fourier transforms of the electronic structure image *before* sublattice segregation signalling a $\mathbf{Q} = 0$ nematic state, as Fig. S6A.

B. Copper site segregated image, $Cu(\mathbf{r})$, in which spatial average is subracted, with copper sites selected from A.

C. x-bond oxygen site segregated image, $O_x(\mathbf{r})$, in which the spatial average is subracted, with x-oxygen sites selected from A.

D. y-bond oxygen site segregated image, $O_y(\mathbf{r})$, in which the spatial average is subracted, with y-oxygen sites selected from A.

**Figure S9 Sublattice Phase Resolved Fourier Analysis for NaCCOC**

A. Measured $\text{Re}\tilde{Cu}(\mathbf{q})$ for NaCCOC sample with $p \sim 12 \pm 1\%$. No DW peaks are discernable at $\mathbf{Q} = (Q, 0), (0, Q)$ or at Bragg satellites surrounding ($\pm$1,0) and (0,$\pm$1). This indicates that the DW in NaCCOC has, like BSCCO, a very small $s$ wave component in its form factor.

B. Measured $\text{Im}\tilde{Cu}(\mathbf{q})$ which also indicates a very small $s$ wave component.

C. Measured $\text{Re}\tilde{O}_x(\mathbf{q})$ showing DW peaks at $\mathbf{Q} = (Q, 0), (0, Q)$ and corresponding Bragg satellites.

D. Measured $\text{Im}\tilde{O}_x(\mathbf{q})$ which exhibits the same structure as C. The colour variation within the DW peaks is smaller for NaCCOC than for BSCCO indicating a less disordered DW.

E. Measured $\text{Re}\tilde{O}_y(\mathbf{q})$ which also shows DW peaks at $\mathbf{Q} = (Q, 0), (0, Q)$ along with Bragg satellites.

F. Measured $\text{Im}\tilde{O}_y(\mathbf{q})$ which exhibits the same structure as E.

**Figure S10 Comparison of $Z(r, E = 150 meV)$ between BSCCO and NaCCOC**

A. Measured $\tilde{O}_x(\mathbf{q})$ for BSCCO sample with $p \sim 8\pm1\%$ obtained using $Z(\mathbf{r}, |E|) = g(\mathbf{r}, E)/g(\mathbf{r}, -E)$, E=150meV.

B. Measured $\tilde{O}_y(\mathbf{q})$ for BSCCO sample using same analysis as in A.

C. Measured $\text{Re}\tilde{O}_x(\mathbf{q}) + \text{Re}\tilde{O}_y(\mathbf{q})$ from A, B. The absence of the four DW peaks at $\mathbf{Q}$ is characteristic of a $d$-form factor DW.

D. Measured $\text{Re}\tilde{O}_x(\mathbf{q}) - \text{Re}\tilde{O}_y(\mathbf{q})$ from A, B. The presence of the four DW peaks at $\mathbf{Q}$ and absence of the Bragg satellite peaks is another expectation for a $d$-form factor DW.

E. Measured $\tilde{O}_x(\mathbf{q})$ for NaCCOC sample with $p \sim 12\pm1\%$ obtained using $Z(\mathbf{r}, |E|) = g(\mathbf{r}, E)/g(\mathbf{r}, -E)$, E=150meV.

F. Measured $\tilde{O}_y(\mathbf{q})$ for NaCCOC sample using same analysis as in E.

G. Measured $\text{Re}\tilde{O}_x(\mathbf{q}) + \text{Re}\tilde{O}_y(\mathbf{q})$ from E, F. The same key signature of a $d$-form factor DW is present in this measurement of NaCCOC as is present in that for BSCCO in C.

H. Measured $\text{Re}\tilde{O}_x(\mathbf{q}) - \text{Re}\tilde{O}_y(\mathbf{q})$ from E, F. The signatures of a $d$-form factor DW are once again seen for NaCCOC in this image and should be compared to that for BSCCO in D.

FIG S1

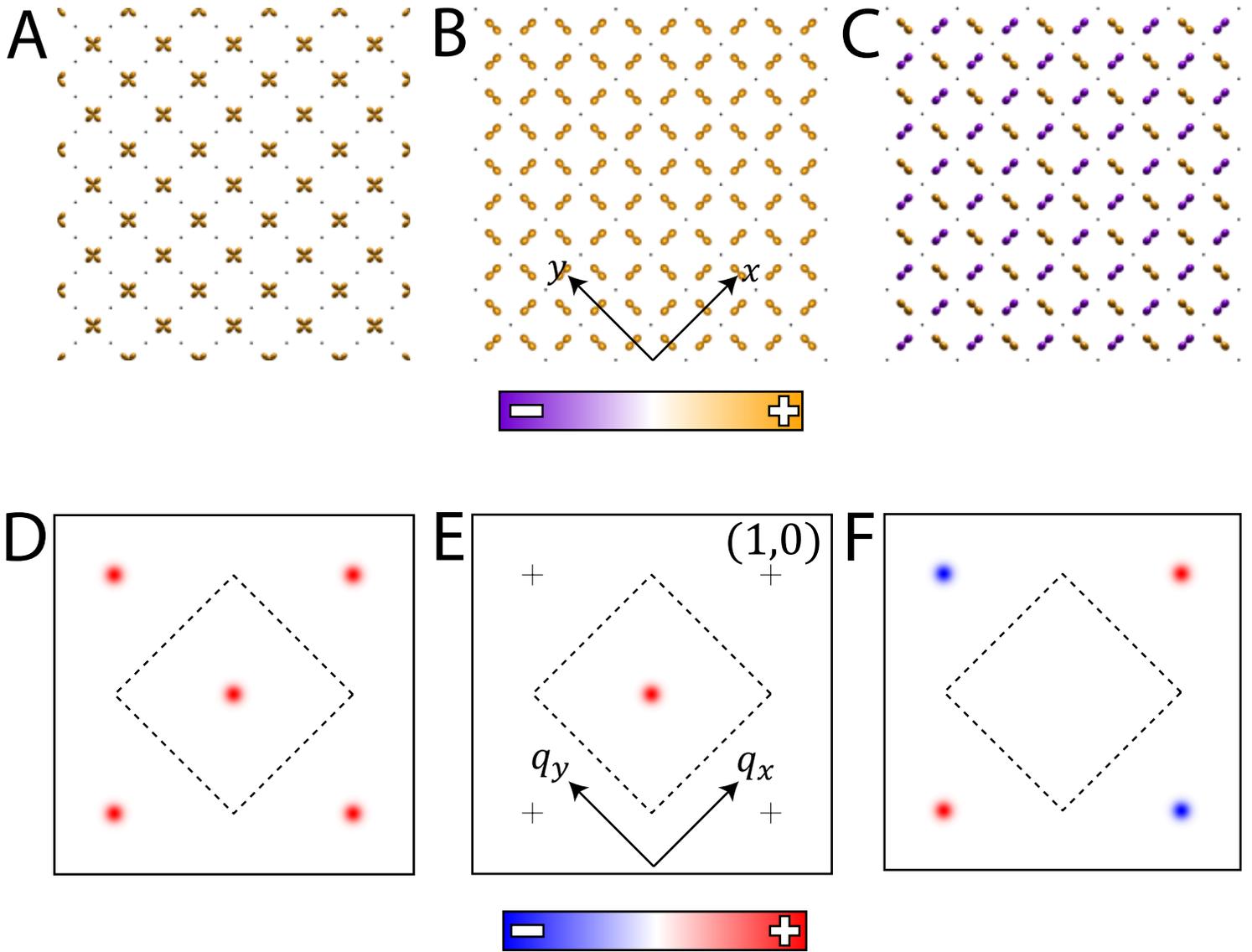

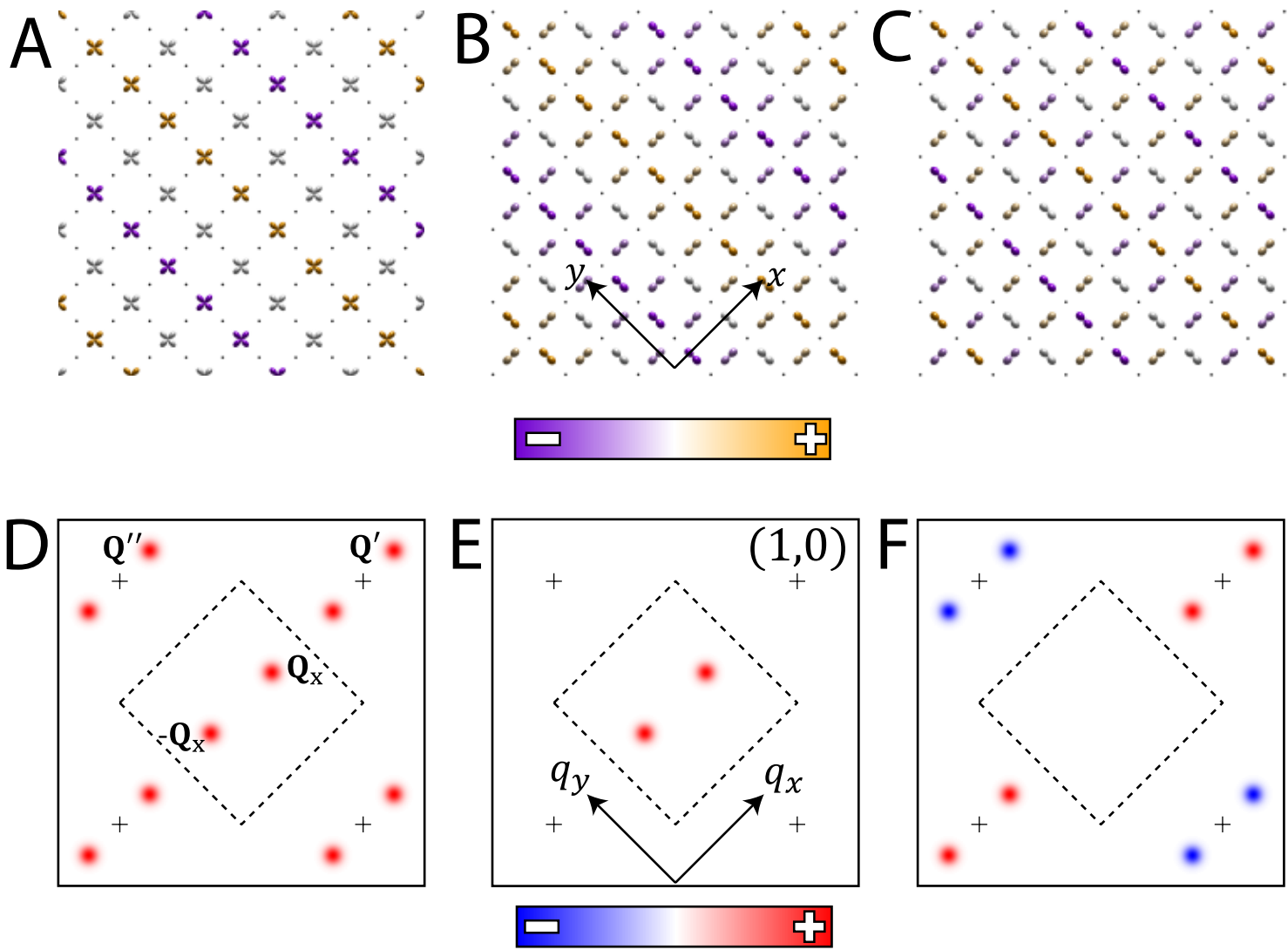

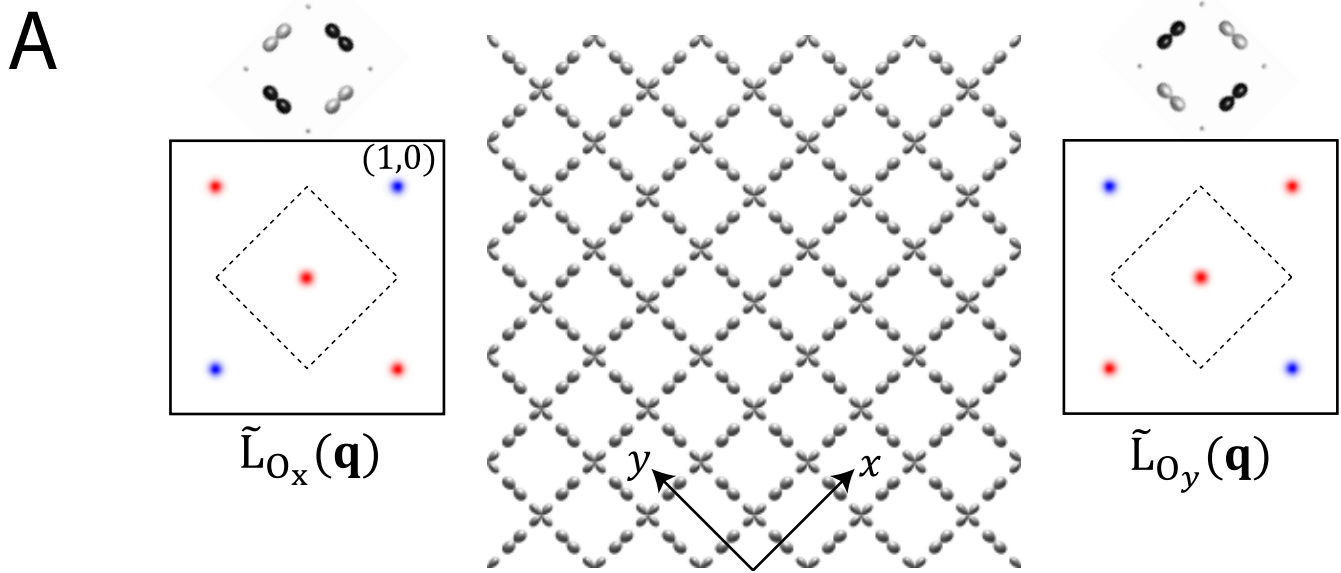

A

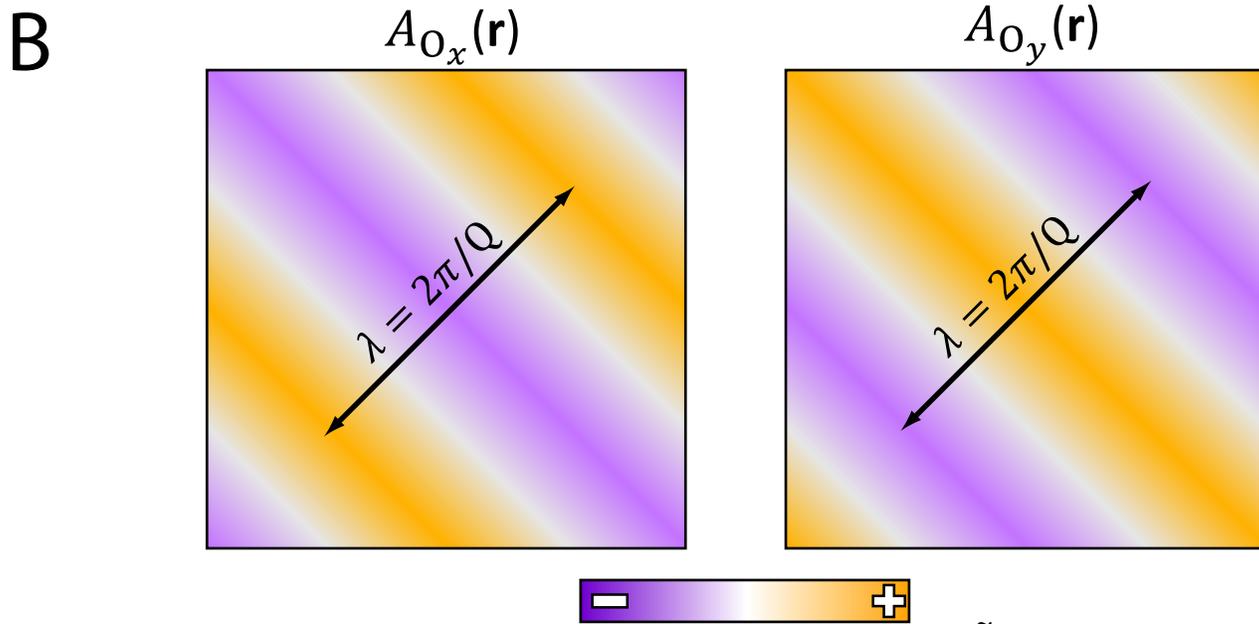

B

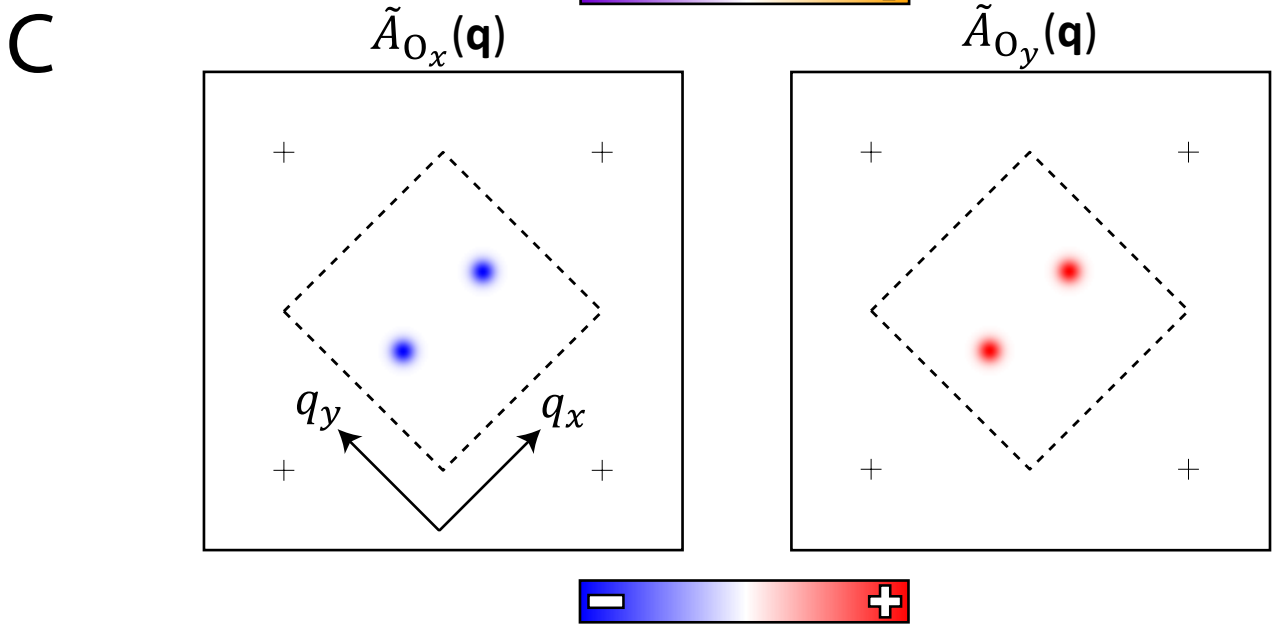

C

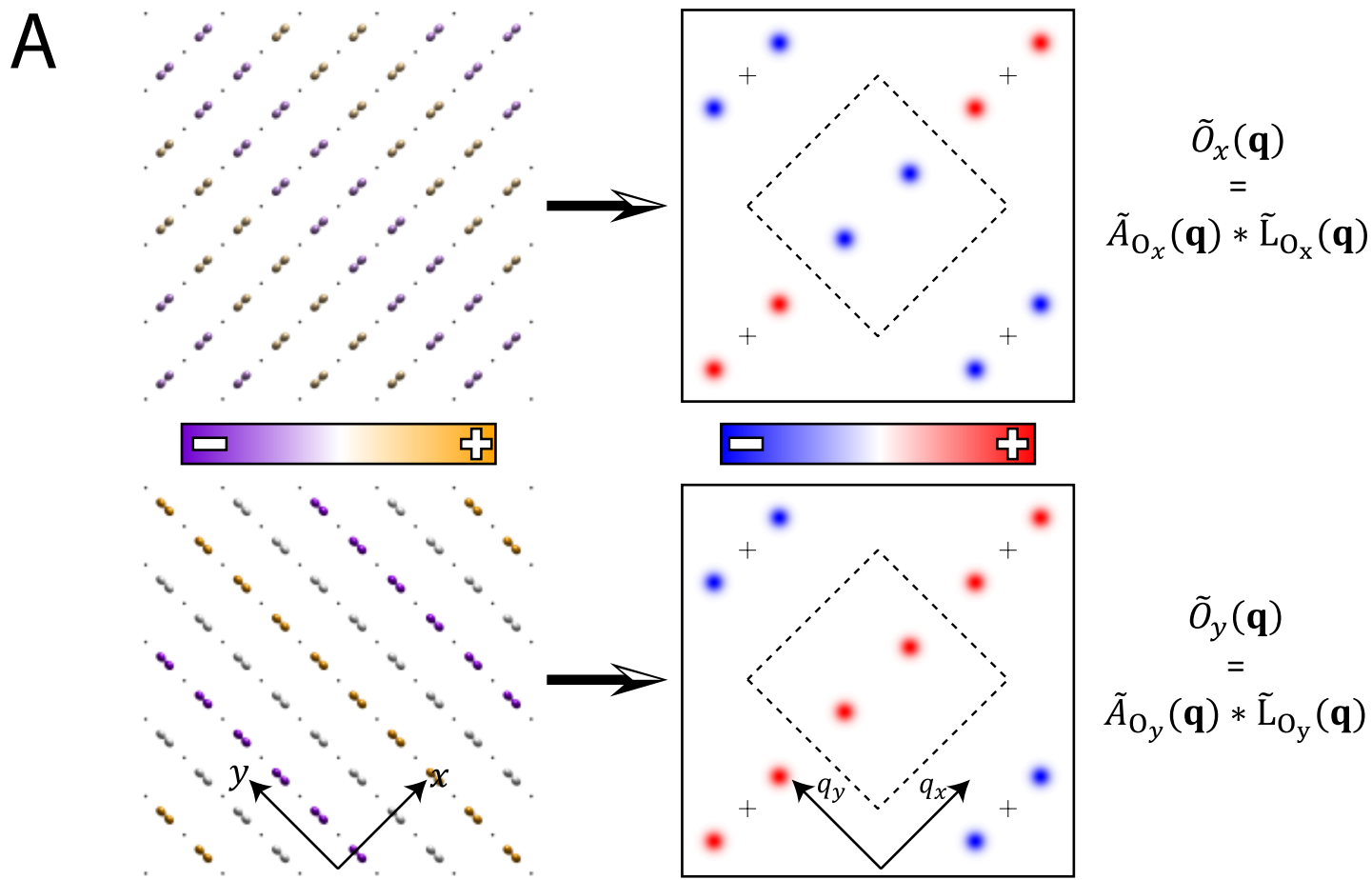

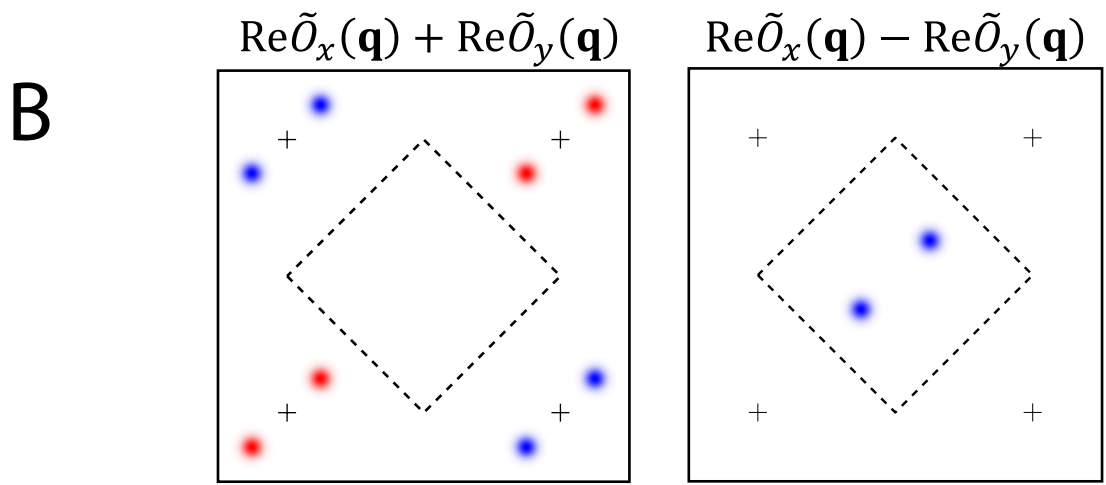

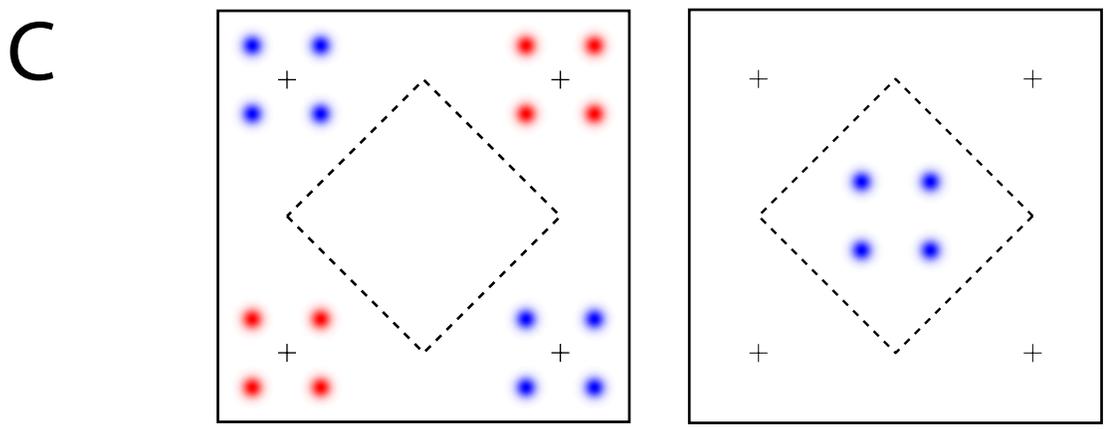

FIG S5

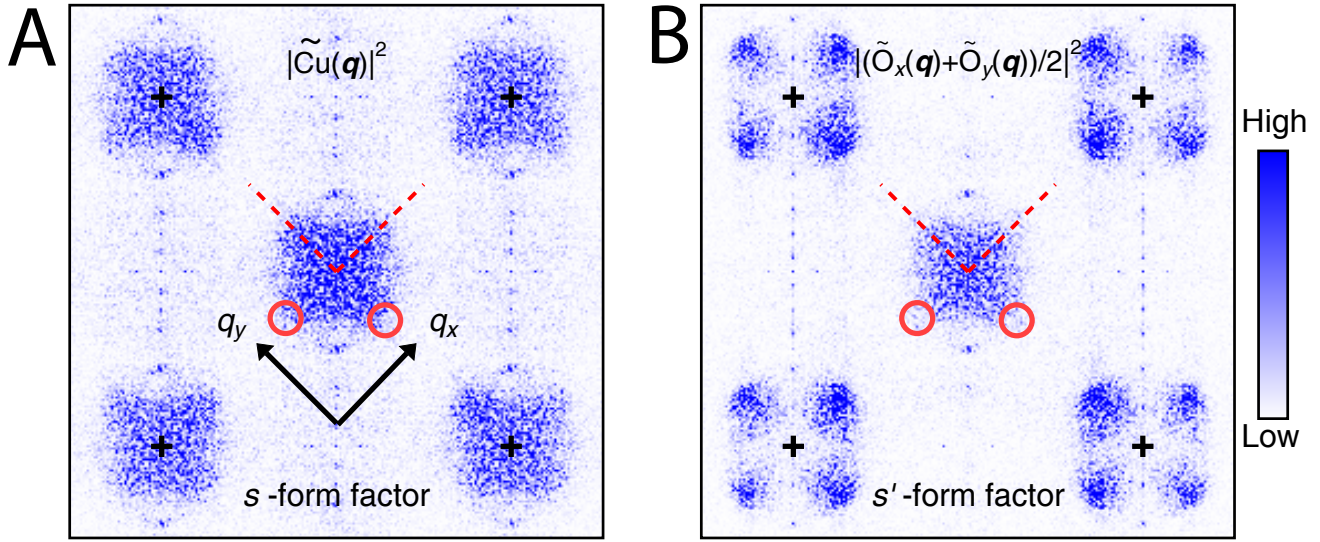

FIG S6

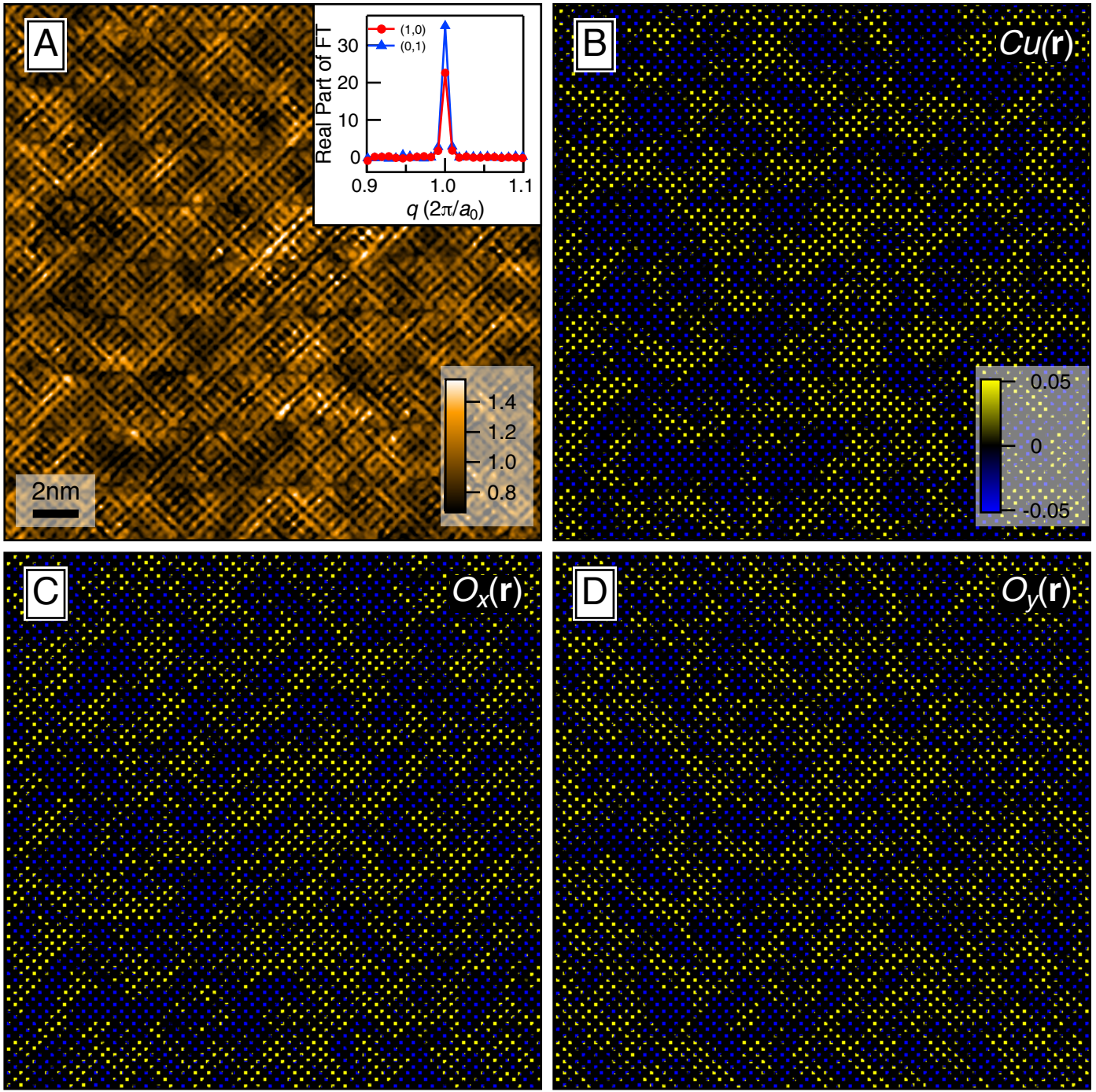

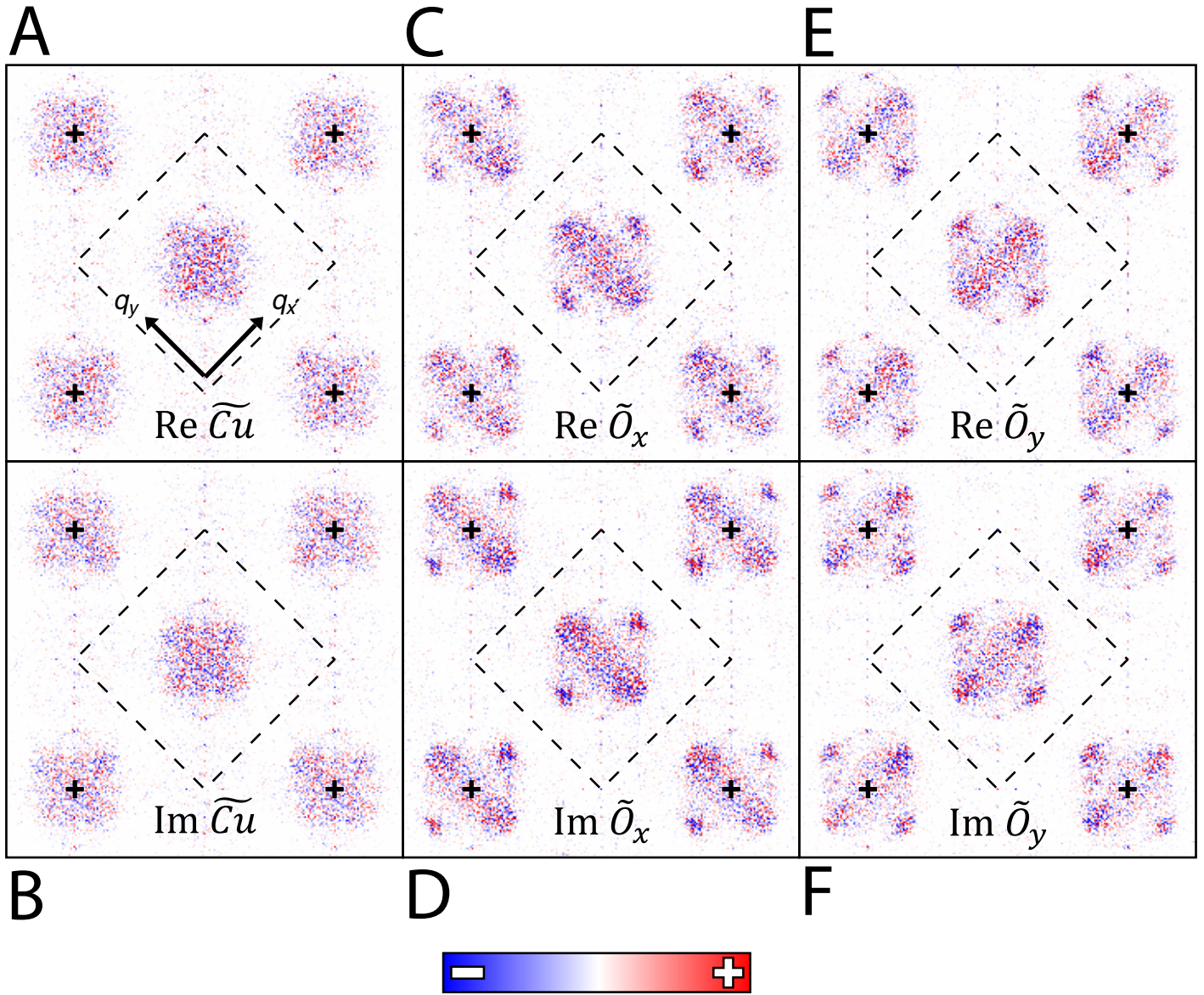

FIG S8

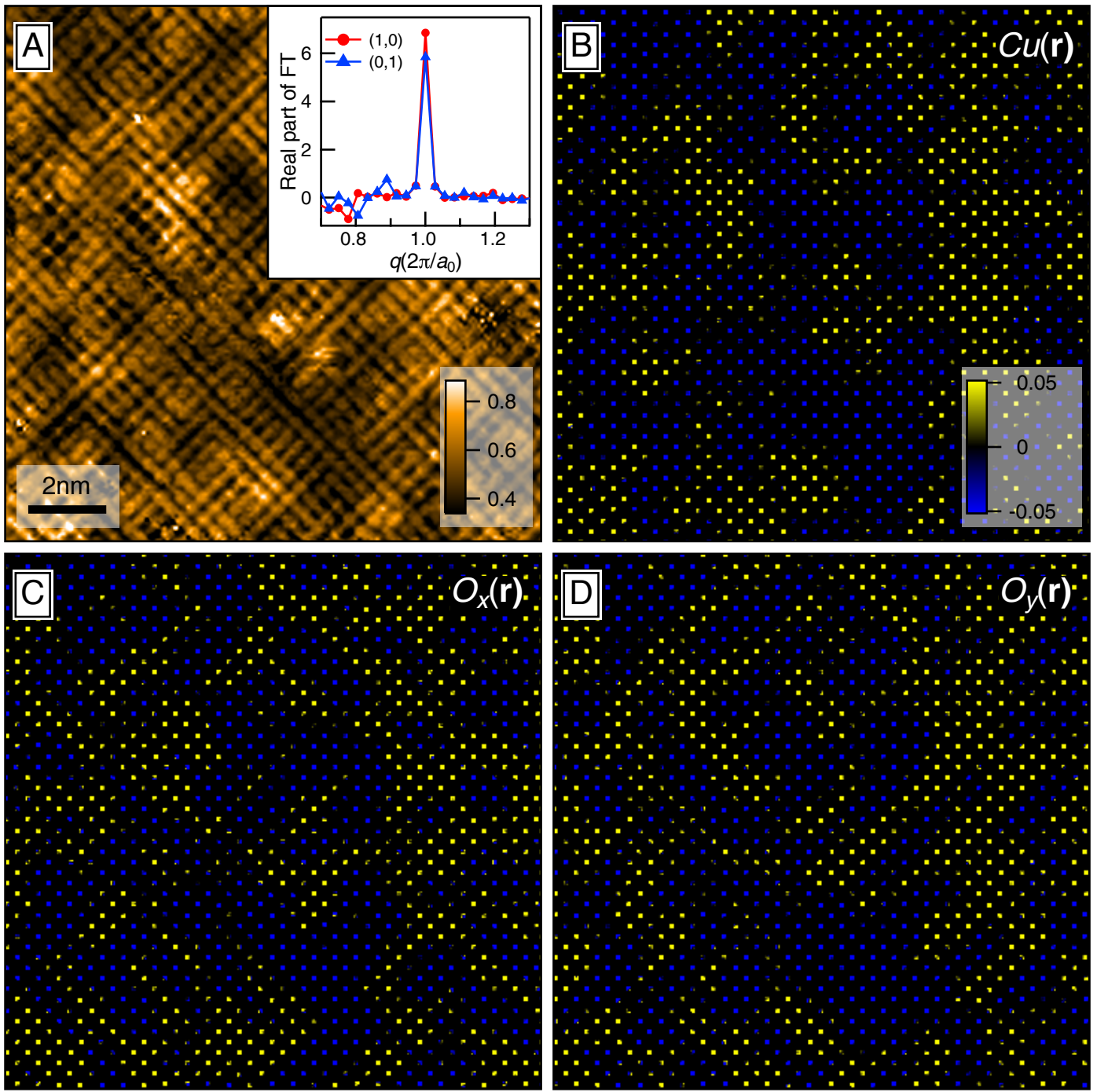

FIG S9

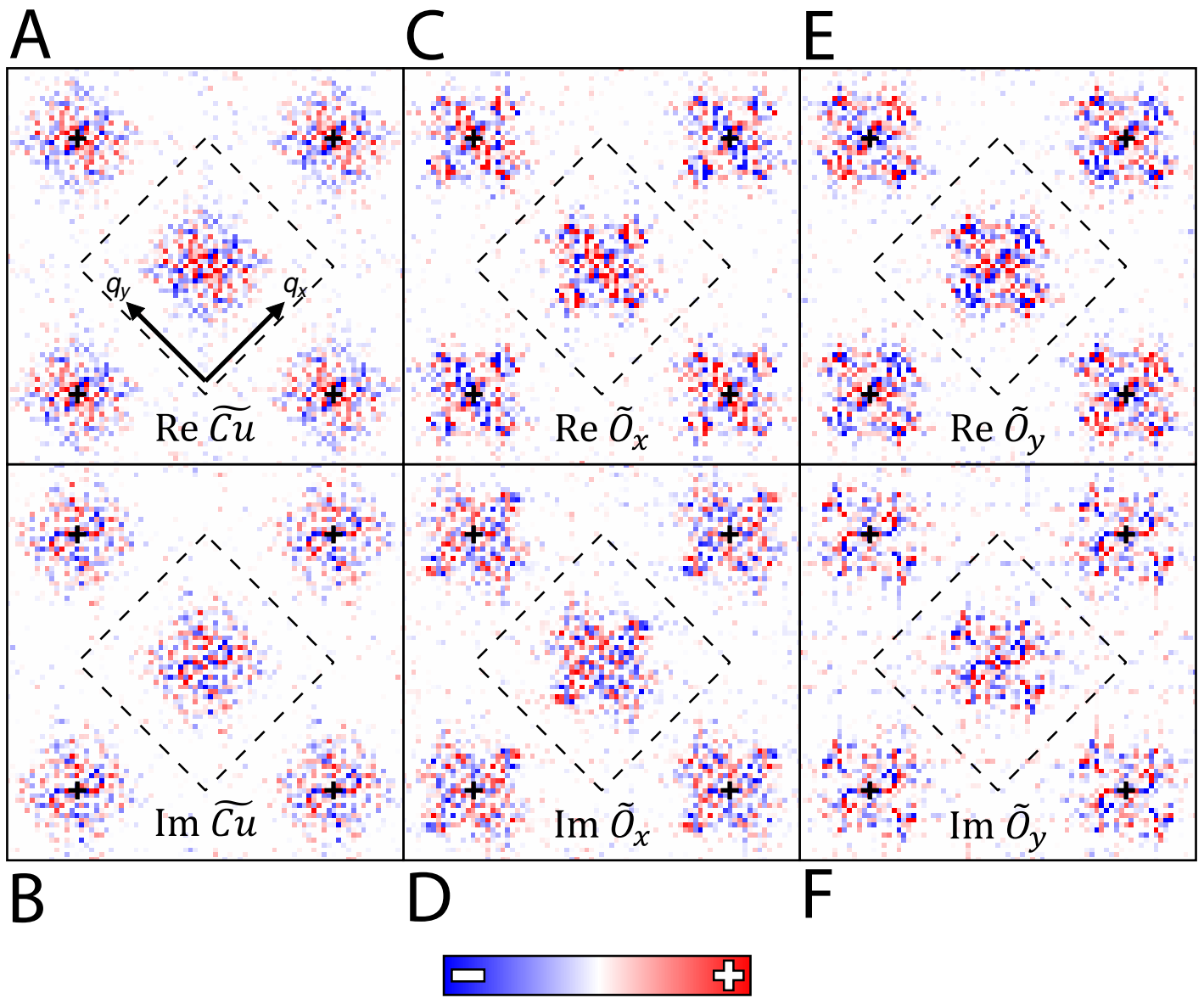

FIG S10

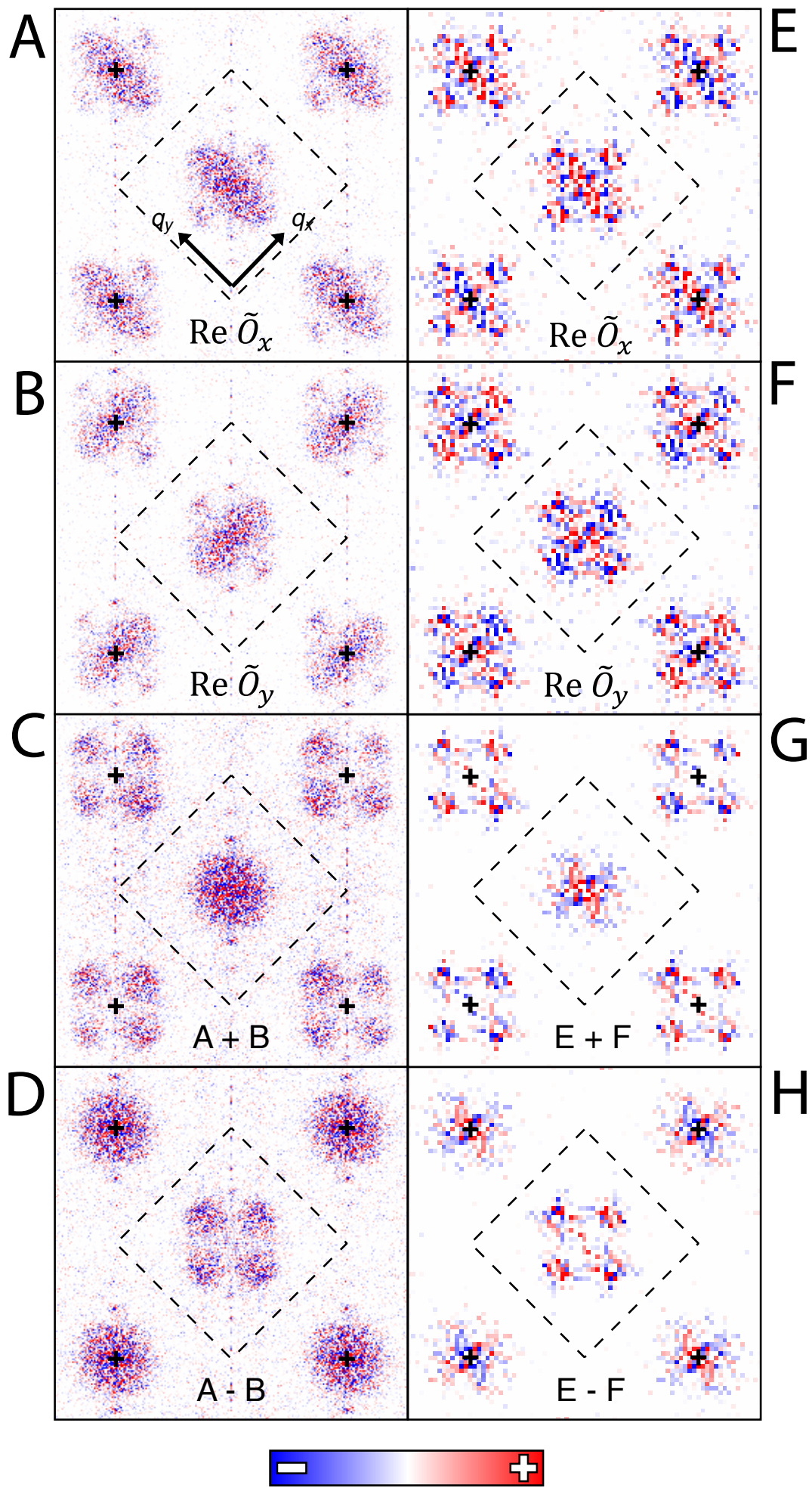